\documentclass[fleqn,usenatbib]{mnras}

\usepackage{newtxtext,newtxmath}

\usepackage[T1]{fontenc}

\DeclareRobustCommand{\VAN}[3]{#2}
\let\VANthebibliography\thebibliography
\def\thebibliography{\DeclareRobustCommand{\VAN}[3]{##3}\VANthebibliography}

\usepackage{graphicx}	
\usepackage{amsmath}	
\usepackage[capitalise]{cleveref} 

\title[Star formation along the early Hubble sequence]{COSMOS-Web: Star formation along the early Hubble sequence and the evolution of dust over the redshift range 0<z<12}

\author[S. A. Eales et al.]{
Stephen Eales,$^{1}$\thanks{E-mail: steve.a.eales@gmail.com}
Matthew W. L. Smith,$^{1}$
Tom Bakx,$^{2}$
Jordan C.J. D'Silva,$^{3}$
Feng-Yuan Frey Liu,$^{4}$\newauthor
and Aparna Venkateshwaran$^{1}$
\\
$^{1}$Cardiff Hub for Astrophysics Research and Technology (CHART), School of Physics and Astronomy,
Cardiff University, 5 The Parade, Cardiff CF24 3AA, UK\\
$^{2}$Department of Space, Earth and Environment, Chalmers University of Technology,
Chalmersplatsen, SE-4 412 96 Gotherburg, Sweden\\
$^{3}$ International Centre for Radio Astronomy Research (ICRAR), University of Western Australia, Crawley, WA 6009, Australia\\
$^{4}$ Institute for Astronomy, Royal Observatory of Edinburgh, 
Blackford Hill, Edinburgh EH9 3HJ\\
}

\date{Accepted 18th May 2026. Received 13 May 2026; in original form 12 Dec 2025}

\pubyear{\the\year{}}

\begin{document}
\label{firstpage}
\pagerange{\pageref{firstpage}--\pageref{lastpage}}
\maketitle

\begin{abstract}
We have carried out a stacking analysis with the COSMOS-Web catalogue on one of the deepest ever
SCUBA-2 images at 850-$\mu$m, allowing us to estimate the mean submillimetre flux density
for samples of galaxies split by stellar mass and morphological class over the redshift
range $0<z<12$. For all morphological classes, the mean star-formation rate estimated from the
dust emission increases with redshift, reaching a value for the most massive galaxies
($\rm M_* \simeq 10^{11}\ M_{\odot}$) of $\rm \gtrapprox 80\ M_{\odot}\ yr^{-1}$ at $2 < z < 4.5$. In this redshift range, the mean
star-formation rate for these galaxies
falls along the Hubble sequence from $\rm \simeq 280\ M_{\odot}\ yr^{-1}$ for irregular galaxies at one end to $\rm \simeq80\ M_{\odot}\ yr^{-1}$for spheroids at the other end, which shows that
quenching was already happening shortly after the emergence of the Hubble
sequence. The decrease in the star-formation rate for the spheroidal
galaxies can be reproduced with a `starvation' quenching model with
a depletion time of $\rm \simeq10^{8.2}\ years$. We also 
show that the transformation of `submillimetre galaxies' can reproduce
the growth in number-density of massive bulge-dominated and spheroidal galaxies over the redshift range $1.5 <z < 4$.
As a side-project, we have used our stacking results
to show that the ratio of dust mass to stellar mass in galaxies increases with redshift out
to $z \sim 8$ and to determine the relationship between the mean density of dust and redshift in the range $\rm 0 < z <12$.
We show that a chemical evolution model based on the `star-formation history' of
the universe, with a gas outflow rate
equal to the star-formation rate, can explain the monotonic rise in the dust-to-stellar mass ratio and reproduce the relationship between mean dust density and redshift
remarkably accurately.

\end{abstract}

\begin{keywords}
galaxies: evolution -- submillimetre: galaxies -- galaxies: high-redshift
\end{keywords}



\section{Introduction} \label{sec:introduction}

The discovery of the `submillimetre
galaxies' (henceforth SMGs) in the first submillimetre surveys was an initial contribution of 
this new kind of astronomy to our understanding of galaxy evolution
\citep{smail97,hughes98,barger98,eales99}. It is difficult to avoid the conclusion
that SMGs are the ancestors of present-day massive ellipticals
for two reasons. First, the spectra of present-day massive ellipticals
imply that most of their stars were formed $\rm \approx 10-12$ Gyr ago in bursts lasting
$\rm \simeq 5 \times 10^8$ years \citep{thomas2005,thomas2010}, 
and SMGs are the only galaxies that we know of that are at the right epoch with
high enough star-formation rates \citep{lilly2013,scott2002,simopson2017,dud2020}. Second, galaxies in the universe today with stellar masses $\rm >2-3\times10^{11}\ M_{\odot}$
are almost entirely ellipticals \citep{kelvin2014}, and the masses of the SMGs are already so high \citep{simopson2017,eales2024a} that it
is hard to imagine what else they could turn into. 

The surveys with the James Webb Space Telescope (JWST) \citep{casey2023,finkelstein2025,stark2026}
are now discovering galaxies well beyond the epoch of the SMGs ($\rm 2 < z < 6$),
galaxies that must be at an early stage of
evolution simply because of the limited time since the big bang. Submillimetre observations are still vital for studying this even earlier population because without them it is easy to underestimate the rate at which stars are forming in these
galaxies. Attempts to
estimate a galaxy's star-formation rate by fitting population synthesis models
to its UV-to-near-IR spectral energy distribution (SED) tend to fail when
the dust obscuration is too high, even though this method makes
an attempt to correct for the effect of dust.
The SMGs are an extreme example, since the typical visual
extinction from SED-fitting is $\rm \sim 2.5\ mags$ \citep{mckinney2025},
whereas the estimates from submillimetre continuum observations of the
dust are $\rm \sim 400\ mags$ \citep{simopson2017,dud2020,harrington2021}, but
recent work suggests this may be a problem for galaxies
with star-formation rates as low as $\rm 100\ M_{\odot}\ yr^{-1}$ \citep{liu2025}.

If SMGs are destined to become massive elliptical galaxies, they still have a long
way to go. From a chemical point of view, they are already at an advanced
evolutionary stage because, based on observations of the dust, CO and [C\textsc{i}], a typical SMG at $\rm z \sim 4$ already contains a mass of metals equal to the entire metal content (in the gas and stars) of a massive early-type
galaxy today \citep{eales2024a}. But since they have very high star-formation
rates and either irregular or disk-dominated structures \citep{Chen2022,Cheng2023,gillman2024,mckinney2025}, they
have undergone neither the
`quenching' of the star formation nor the morphological transformation
necessary to turn them into a massive elliptical galaxy.

The causes of morphological transformation and quenching are, of course,
two of the big unknowns in our understanding of galaxy evolution in general and
are two of the themes of this paper.
The two main processes that have been suggested for creating the structures
of elliptical galaxies are galaxy merging \citep{toomre1972,hopkins2008} 
or the collapse of gas with low angular momentum \citep{granato2004,ubler2014}. The processes suggested as the cause of the quenching are more varied, including
a merger-triggered starburst \citep{hopkins2008}, radiative-mode \citep{fabian2012,bollati2024} or radio-mode \citep{croton2006,dubois2016}
AGN feedback, some other `starvation'/`strangulation' process \citep{larson1980,peng2015}, stabilisation of the gas in a disk by the growth of
a bulge \citep{martig2009}, or some environmental process \citep{peng2010} that strips away
the gas such as ram-pressure stripping. In this paper (as in the literature), we often casually refer to both the reduction in the star-formation rate or the removal of a galaxy's gas
as `quenching', since for most of the processes above these are the same thing, but it is important to note that if disk-stabilisation \citep{martig2009} is the cause,
a galaxy's gas reservoir may stay full even though the star-formation rate drops.

The morphological transformation and
quenching do not necessarily happen at the same time or on
the same timescale. In principle, it is possible
to observationally separate these two processes, because different
combinations in the lists above make different predictions about the temporal relation between the morphological transformation and the quenching.

Suppose, for example, that the
emergence of an elliptical galaxy is the result of the merging of two galaxies
with the merger triggering
a starburst that rapidly depletes the gas \citep{hopkins2008}. When we look back to the epoch
at which ellipticals emerge, we would expect to find them already with low star-formation rates and largely depleted of gas. 
This would also be the case if the infall of gas
into the centre of the new galaxy as the result of the merger led to
the formation of an AGN, which then expelled the gas through radiative-mode
feedback \citep{hopkins2008}.
On the other hand, if the morphological transformation was accomplished by
merging but the quenching occurred later more slowly as the result of starvation by
radio-mode feedback or some other means \citep{larson1980,croton2006,peng2015}, we would expect to find high-redshift ellipticals that still have high star-formation
rates and large gas reservoirs.

Before the era of JWST and {\it Euclid}, this simple test was
hard to perform because of the limited information about the morphologies
of high-redshift galaxies, largely confined to the morphologies measured
with the {\it Hubble} CANDELS survey \citep{wuyts2011}.
In those days, the standard approach was to try to measure the gas
content of high-redshift quiescent galaxies, on the
assumption that these are mostly early-type galaxies, for which there
was fairly good evidence \citep{wuyts2011,gobat2018}. The standard technique
used to estimate the gas mass of a galaxy was to measure the dust mass, and so
infer the gas mass,
since dust has major advantages over CO for tracing the molecular phase
in galaxies \citep{magdis2012,eales2012,scoville2014,tacconi2020,dunne2022}. This approach
had contradictory results, with some statistical studies \citep{gobat2018,magdis2021} finding
that the dust masses of massive quiescent galaxies are higher at $\rm z \sim 1-2$ than
they are today, although
not as high as those of star-forming galaxies at the same redshift, and other studies
\citep{whitaker2021} finding much lower dust masses.

The advent of JWST and {\it Euclid} has changed this field completely, making it possible
to classify the morphologies of galaxies over large areas of the sky \citep{quilley2025,huertas2025}.
These studies are rapidly advancing our understanding of galaxy evolution. Observations with {\it Euclid} have already
shown that the common assumption of there being two basic classes of galaxies - quiescent/early-type and
star-forming/late-type - is too simplistic, since in 
a study of the morphologies of nearly a million
galaxies in the redshift range $\rm 0.25 < z < 1$, only 82\% fell in the standard categories, with 13\% of the galaxies consisting of star-forming spheroids and 5\% of quiescent disks \citep{gentile2025}.
Observations with JWST have shown that the familiar morphologies
of the Hubble sequence seem to emerge at $\rm z \sim4$, with galaxies
at higher redshift either being unresolved or classified as irregular \citep{huertas2025}.

In this paper, we use morphological classifications
from the JWST COSMOS-Web survey \citep{casey2023,shuntov2025a} to make measurements of the submillimetre emission from dust
for samples of galaxies split by stellar mass, redshift and, for the first
time, morphological class, which allows us to directly address the issue of whether morphological transformation is linked to the quenching of the star formation.
Although it is possible to measure the submillimetre emission from individual galaxies
with interferometers such as the Atacama Large Millimeter/Submillimeter Array (ALMA),
it is very challenging to do this for large numbers of galaxies. For this
reason, we have used one of the deepest ever submillimetre survey carried out with a single-dish telescope, the James Clerk Maxwell
Telescope (JCMT), mitigating the effect of the poor resolution (relative to the JWST and {\it Euclid}) by carrying out a `stacking analysis' in which we measure the mean submillimetre flux density for samples of galaxies rather than individual galaxies.

As a second science project, we have also used our stacking results to investigate the general
properties of dust in the universe.
This paper is a follow-up to \citet{eales2024b}, in which we used a submillimetre stacking analysis for the galaxies in the COSMOS survey \citep{scoville2007} to determine how the mean density of dust in the universe depends on redshift out to $\rm z \sim 5.5$. 
Although COSMOS-Web doesn't cover as large a field of view as COSMOS ($\sim$0.54\ deg$^2$), it still is the largest JWST survey
and thus the one least affected by cosmic variance. The release of its
data \citep{shuntov2025a} has allowed us
to extend our investigation of dust in the universe out to  $\rm z \sim 12$.

The paper is organized as follows: \cref{sec:methods} describes our methods; \cref{sec:results} gives our results; \cref{sec:discussion} is the discussion; our results and conclusions are summarized in \S5. Given that this is such
a long paper, we note that for those primarily interested in galaxy evolution
the key results are in sections 4.1, 4.2, 4.3, 4.4 and 4.6. For those primarily
interested in the evolution of dust, the key results are in sections 3.4, 3.5
and 4.5.

We use the cosmological parameters given in \citet{planck2015}.

\begin{figure*}
    \includegraphics[width=\textwidth, clip=True, trim=0mm 21mm 0mm 140mm]{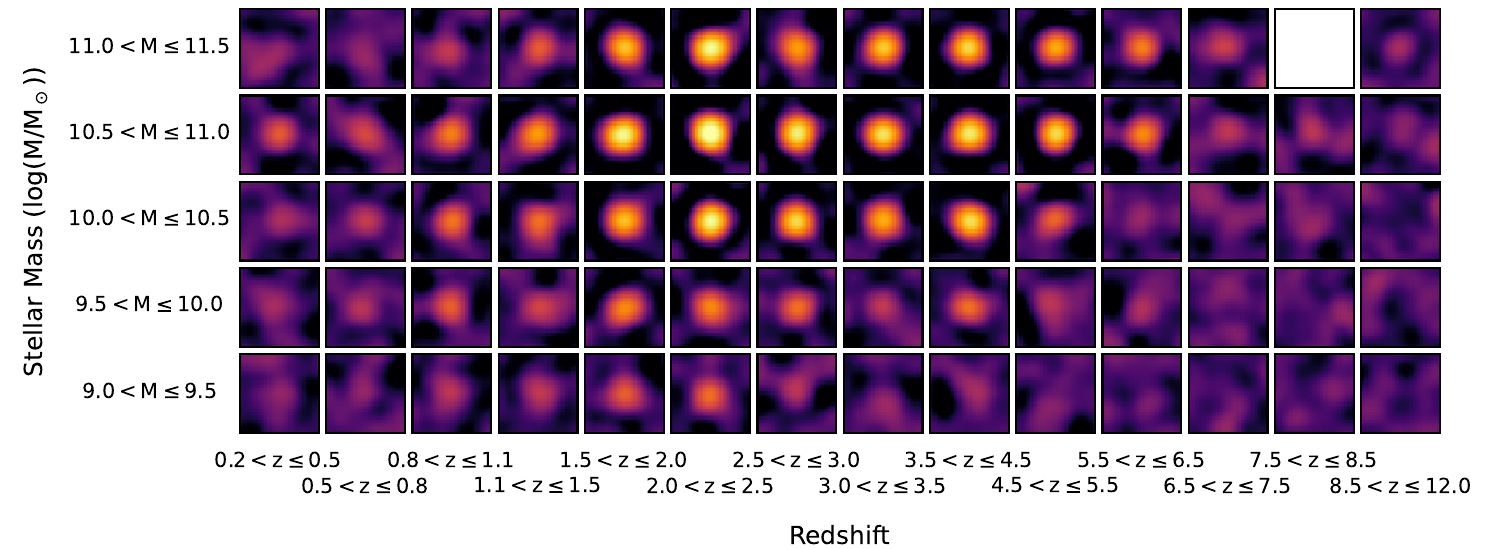}
    \includegraphics[width=\textwidth, clip=True, trim=0mm 21mm 0mm 140mm]{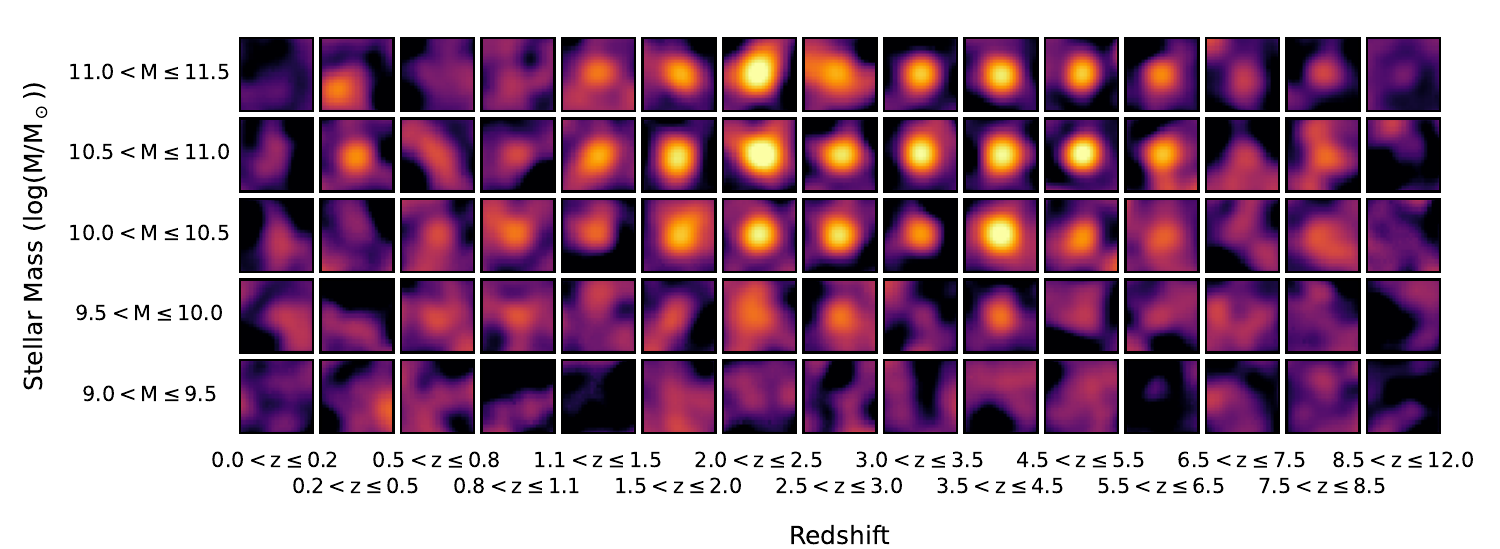}
	\caption{Images showing the signal-to-noise of our measurements
    of the mean 850-$\mu$m flux density of 70 samples of galaxies, with the results
    from the basic stacking method shown above and from SIMSTACK shown below. The annotation shows the
    redshift and mass range for each sample. Each pixel in each image represents a different offset
    from the COSMOS-Web positions, with the zero-offset position being in the centre. The images are squares with
    a side of $44\arcsec$ and a pixel size of $2\arcsec$. The blank field in the top
    panel is for a sample containing less than four galaxies, the minimum number
    for our basic stacking method.}
	\label{fig:dust density}
\end{figure*}

\section{Methods} \label{sec:methods}

\subsection{Overview} \label{subsec:methods_overview}

The basic technique used in this paper is `stacking'.
In a stacking analysis, measurements of the flux density of
a sample of galaxies are used to calculate the weighted-mean flux
density for the sample, which 
is a very useful technique when
many of the individual measurements have low signal-to-noise.
In our study, we have carried out a stacking analysis using the images from the 
S2COSMOS 850-$\mu$m survey of
the COSMOS field \citep{simpson2019}.

\subsection{The submillimetre image}

The 850-$\mu$m image was constructed using 640 hours of new and archival
observations taken with the SCUBA-2 camera on the James Clerk Maxwell 
Telescope as part of the SCUBA-2 COSMOS survey (S2COSMOS). It achieved
a median noise level of $\sigma_{850\ \mu m} = 1.2\ mJy\ beam^{-1}$ over
1.6 square degrees which covers the COSMOS-Web region \citep{simpson2019}. The angular
resolution of the image is $\simeq$15 arcsec (FWHM).

\subsection{The COSMOS-Web samples} \label{subsec:methods_samples}

Our samples were taken from the COSMOS-Web catalogue\footnote{Public data release: \href{https://cosmos2025.iap.fr}{COSMOS-Web DR1}} described by \citet{shuntov2025a}. We only used
objects from the catalogue if they had $\it warn\_flag = 0$\footnote{A detailed description of how the data was flagged is included in the public data release page.}, removing problematic objects such as those suspected of being hot pixels, or those with inconsistencies between the JWST and ground-based photometry.

In our analysis, we used the estimates of the photometric redshift ($\tt zfinal$),
stellar mass ($\tt mass\_med$) and specific star-formation rate 
($\tt ssfr\_med$) in the COSMOS-Web catalogue, which were
obtained with the LePHARE (Photometric Analysis for Redshift Estimate) code for fitting spectral energy distributions \citep{arnouts99}, which is based on models of the stellar population of a galaxy \citep{bruzual2003}. We restricted our analysis to galaxies with stellar masses
in the range $9.0 < log_{10}M_* < 11.5$ and redshifts in
the range $0.2 < z < 12$. We chose the lower mass limit because this is approximate stellar mass completeness limit of COSMOS-Web \citep{shuntov2025a}\footnote{COSMOS Web is $\simeq80\%$ complete to this stellar mass at $z\sim10$ \citep{shuntov2025b}}.
We chose the upper mass limit because high-redshift galaxies at higher masses are likely to be spurious, the
effect of Eddington bias \citep{casey2024}. To be consistent with estimates
of the COSMOS-Web stellar mass function, we removed objects at $z>3.5$
for which there is evidence that they are either AGN or `little red dots' using the same selection
criteria as used in calculating the stellar mass function
\citep{shuntov2025a}. After removing these, our base sample consisted of 102,549 galaxies.

In our initial analysis, we divided the galaxies into 14 redshift intervals (\cref{tab:sample_redshift_range}) over the redshift range $0 < z < 12$ and five equal intervals
of stellar mass ($M_*$) over the range $\rm 9.0 < log_{10}\ M_* < 11.5$.
This gave us 70 different samples of galaxies, the goal of the stacking
being to produce an unbiased estimate of the mean 850-$\mu$m flux density of each sample.

At a later stage in the analysis, we also divided the galaxies into morphological types. We used the morphological classifications given in the ML-MORPHO extension of the COSMOS-Web catalogue.
These were obtained by using a machine-learning algorithm trained on the {\it Hubble} CANDELS survey to
classify the COSMOS-Web galaxies into four broad morphological classes: spheroids, bulge-dominated, disk-dominated
and irregular \citep{huertas2025}.
The catalogue includes three classifications for each galaxy
made using images made through three JWST filters. We followed \citet{huertas2025} in 
choosing the morphological classification for each galaxy so they were all
done at roughly the same rest-frame wavelength. We used the classifications
from the F150W, F277W and F444W images for the redshift ranges
$\rm z < 1$, $\rm 1 < z <3$, and $\rm z > 3$, respectively.
Of the sample, $\simeq97\%$ of the galaxies have morphological classifications.
The morphological breakdown of each of our 70 samples is given in Appendix A. Although the overall percentage with no morphological
classification is very small, it does reach very high values in some
of the high-redshift bins. Our investigation of the relationship between star-formation rate and morphological type (Sections 4.1 to 4.4) is therefore restricted to $z < 4.5$. The largest percentage
incompleteness below this redshift is 35\% for the sample with
$3.5<z<4.5$ and $11.0 < log_{10}(M_*) < 11.5$.

\subsection{Stacking - general points} \label{subsec:methods_stacking_general_points}

Although the relatively poor
angular resolution of the SCUBA-2 image (FWHM of 15\arcsec) means there are many
JWST galaxies within each resolution element, stacking will produce an unbiased estimate
of the mean flux density of a sample as long as the mean of the image is zero
and there is no clustering between galaxies \citep{marsden2009,viero2012}. Two things complicate matters. First, obviously, galaxies are clustered. Second, the redshifts in the
COSMOS-Web catalogue are photometric redshifts, which means the galaxies are not necessarily at their
correct redshifts.

We attempted to solve the first problem by using the SIMSTACK algorithm \citep{viero2013},
which is an elegant way of correcting for the effect on the measurements of clustering between
galaxies in different samples. We attempted to correct for the second problem by using a Monte-Carlo simulation
to estimate the additional errors
in the stacked flux densities arising from errors in the redshifts and stellar masses. We generated
100 artificial COSMOS-Web catalogues by using the redshifts and stellar masses in the real catalogue, and the errors on both, to generate bootstrapped sets of stellar masses and redshifts for each
artificial catalogue.

We carried out two versions of the stacking analysis. The first version used
the basic method, in which we made no attempt to correct for clustering. The second version
was the SIMSTACK version. Our reason for doing so was that the results from the basic
method gave us a useful check on the results from SIMSTACK. The differences between the two sets of results
also gave us insight into the actual effect of clustering
on stacking. In the following two subsections we give some technical details of how we applied the two
methods.

\begin{table}
	\centering
	\caption{Redshift ranges of the samples and results from fitting equation 5}
	\label{tab:sample_redshift_range}
	\begin{tabular}{lcc} 
		\hline
		Redshift ranges & a & b\\
		\hline
		$\rm 0.2<z<0.5$ & $\rm -7.1\times10^{-4}$  & $\rm 8.2\times 10^{-3}$\\
        $\rm 0.5<z<0.8$ & $\rm -1.7\times10^{-3}$  & $\rm 1.9\times 10^{-2}$\\
        $\rm 0.8<z<1.1$ & $\rm -1.7\times10^{-3}$  & $\rm 2.0\times 10^{-2}$\\
        $\rm 1.1<z<1.5$ & $\rm -1.6\times10^{-3}$  & $\rm 1.9\times 10^{-2}$\\
        $\rm 1.5<z<2.0$ & $\rm -2.7\times10^{-3}$  & $\rm 3.1\times 10^{-2}$\\
        $\rm 2.0<z<2.5$ & $\rm -3.7\times10^{-3}$  & $\rm 4.4\times 10^{-2}$\\
        $\rm 2.5<z<3.0$ & $\rm -4.8\times10^{-3}$  & $\rm 5.6\times 10^{-2}$\\
        $\rm 3.0<z<3.5$ & $\rm -3.7\times10^{-3}$  & $\rm 4.5\times 10^{-2}$\\
        $\rm 3.5<z<4.5$ & $\rm -6.6\times10^{-3}$  & $\rm 7.9\times 10^{-2}$\\
        $\rm 4.5<z<5.5$ & $\rm -3.7\times10^{-3}$  & $\rm 5.1\times 10^{-2}$\\
        $\rm 5.5<z<6.5$ & $\rm -5.2\times10^{-3}$  & $\rm 6.8\times 10^{-2}$\\
        $\rm 6.5<z<7.5$ & $\rm -4.5\times10^{-3}$  & $\rm 5.7\times 10^{-2}$\\
        $\rm 7.5<z<8.5$ & $\rm -5.5\times10^{-3}$  & $\rm 7.3\times 10^{-2}$\\
        $\rm 8.5<z<12.0$ & ... & ... \\
		\hline
	\end{tabular}
\end{table}

\subsection{Basic Stacking Method} \label{subsec:methods_basic_stacking}

In this version of the method, we made no attempt to correct for the clustering between
the galaxies in different samples. We simply measured the flux density and noise from the S2COSMOS images for each
of the galaxies in a sample, and then calculated the mean flux density for the sample, with the
weight for each galaxy, $w_i$, given by:

\begin{equation}\label{eq:stacking_weights}
    w_i = {{1 \over \sigma_i^2} \over \sum_i {1 \over \sigma_i^2}} 
\end{equation}

\noindent in which $\sigma_i$ is the noise for the galaxy and the sum 
is over all the galaxies in the sample. In this method, we used the S2COSMOS image and corresponding noise
image that
had been smoothed with a filter equal to the point spread function of
the observations to maximize the signal-to-noise.

To calculate the error on the stacked signal that is as close as possible to ground truth,
we carried out a Monte-Carlo simulation for each sample. If there were $N_{gal}$ galaxies in
the sample, we generated 1000 sets of $N_{\rm gal}$ random positions. We then carried out a stacking analysis
for each set. We used the variance of the 1000 weighted mean values to estimate the error in the stacking
analysis: $\sigma_{\rm stack}$. 

As described above, we also produced 100 artificial versions of the COSMOS-Web catalogue, using the
photometric redshifts and the stellar masses and their associated errors
to generate a set of bootstrapped redshifts and stellar masses for each artificial catalogue. We carried
out the same stacking analysis for each artificial catalogue, using the variance of the results
to calculate the error that arises from the errors in the redshifts and the stellar masses: $\sigma_{\rm JWST}$.
We added $\sigma_{\rm stack}$ and $\sigma_{\rm JWST}$ in quadrature to get the final
error on the stacked signal.

To produce a more visual way of judging the significance of the stacked signal, we generated images that show the signal-to-noise of
the mean 850-$\mu$m flux density of each sample.
Each image is produced from a 23 $\times$ 23 array of offsets from the JWST positions centred on
(0,0) and with a grid-spacing of $2\arcsec$. The image value at each pixel
is the signal-to-noise of the mean 850-$\mu$m flux density produced by
the stacking analysis when the galaxies
are placed at this offset from their true positions. The upper panel of
Figure 1 shows these signal-to-noise images.

\begin{figure*}
	\includegraphics[width=\textwidth]{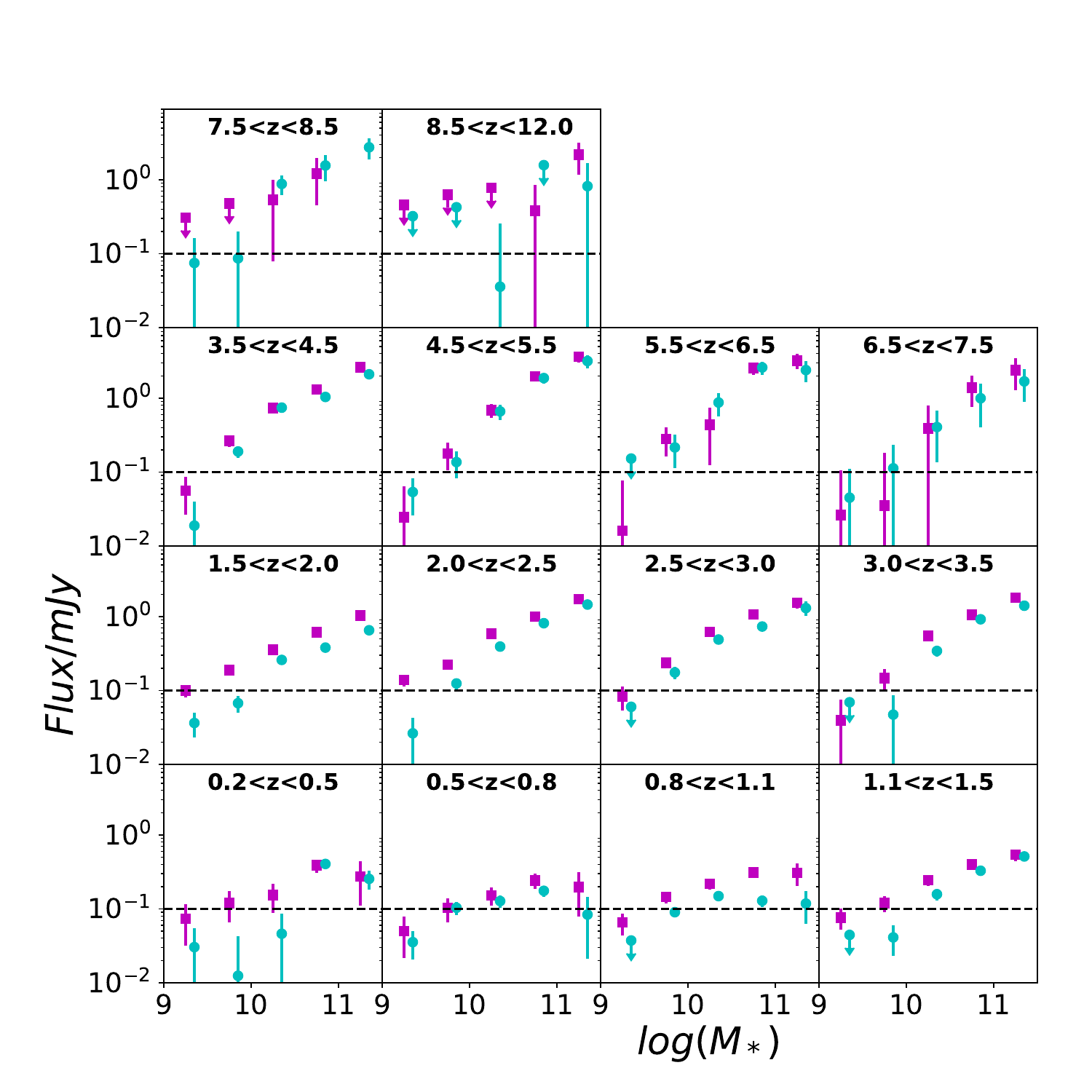}
	\caption{Mean flux density at 850 $\mu$m versus mass for each redshift interval. The mauve points are the values measured with the basic stacking method. The cyan points are the values measured with SIMSTACK. The two sets have been offset by 0.1 in $\rm log_{10}(M_*)$, to prevent an overlap. The horizontal dashed line
    at a flux density of 0.1 mJy has been drawn to aid comparison between
    the panels. The limits are 3$\sigma$ upper limits.}
	\label{fig:dust_density}
\end{figure*}

\subsection{Stacking using SIMSTACK} \label{subsec:methods_stacking_simstack}

In this version, we used the SIMSTACK algorithm \citep{viero2013,viero2022}.
The SIMSTACK algorithm fits the contribution to the submillimetre image
from all samples simultaneously, which means it is not biased by the effect
of clustering between galaxies in different samples.
There is a particularly clear exposition
of the maths of the algorithm in \citet{hill20225}. The implementation of the algorithm that we used in this paper is
the one available at \href{https://github.com/marcoviero/simstack3/tree/main}{SIMSTACK3}.

We followed the same procedure as described in \citet{viero2022}. 
In this case, we used the unconvolved S2COSMOS images.
As the result of the errors in the photometric 
redshifts, there will be a contribution to the stacked signal from clustering
between galaxies that are in samples
in different redshift intervals. We therefore followed \citet{viero2022} in applying SIMSTACK
to samples
at all redshifts simultaneously. We included an additional sample 
in the SIMSTACK analysis that consisted of all the galaxies in the COSMOS-Web catalogue outside
the mass and redshift limits of the other 70 samples. To avoid errors being produced by holes
in the coverage of the COSMOS-Web catalogue, we included a flat foreground layer in SIMSTACK \citep{viero2022}.

As in the other version of the stacking, we applied SIMSTACK to the 100 artificial versions of the COSMOS-Web
catalogue to estimate the error produced from the errors in the photometric redshifts and stellar masses: $\sigma_{JWST}$.
We added this error in quadrature to the error produced by SIMSTACK, $\sigma_{stack}$, to
get the final error on the mean flux density.

As in the other stacking method, we 
produced cutout images showing the signal-to-noise for each sample, which we generated
using the method described in the previous section. These are shown in the lower panel
of Figure 1.

\begin{figure*}
	\includegraphics[width=\textwidth]{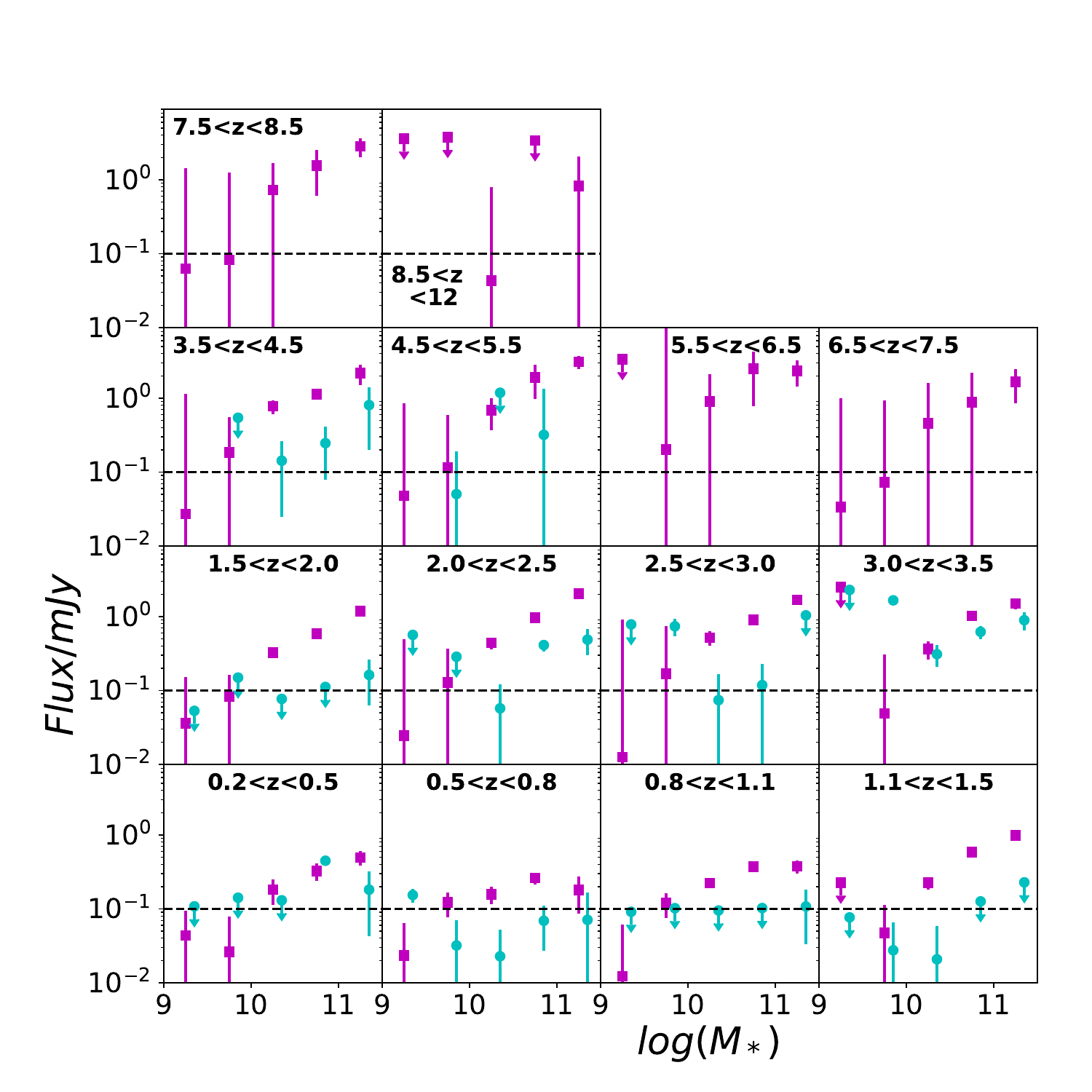}
	\caption{Mean flux density at 850 $\mu$m estimated with SIMSTACK (\S 2.4) for star-forming
    galaxies (mauve) and quiescent galaxies (cyan) plotted against stellar mass for the
    14 redshift intervals. The two sets of points have
    been offset by 0.1 in $\rm log_{10}(M_*)$ so that they
    don't overlap. The horizontal dashed
    line at a flux density of 0.1 mJy has been drawn to
    aid comparison between the panels. The upper limits
    are 3$\sigma$ upper limits.}
	\label{fig:dust density}
\end{figure*}

\subsection{The Temperature Models}

Both the stacking methods above produce an estimate of the mean submillimetre flux density at a single
wavelength. They do not produce an estimate of the temperature of the dust, which is essential for 
an accurate estimate of the mass of dust. In a similar way, it is only possible
to estimate the total (bolometric) emission of the dust, which can be
used to estimate the galaxy's star-formation rate \citep{kennicutt1998,kennicutt2012}, if one knows the temperature of the dust or, equivalently, has a model of
the SED of the dust emission. The accuracy of the estimates of both, of course,
depend critically on the accuracy of the assumptions about the dust temperature
or the SED.

There is no consensus about whether the temperature of the dust in galaxies does generally increase with
redshift, with evidence both for 
\citep{magdis2012,magnelli2014,bethermin2015,ivison2016,
faisst2017,schreiber2018,Bouwens2020,viero2022,witstok2023} and against this \citep{swinbank2014,casey2018,jin2019,dud2020,drew2022,bendo2023,ward2024}. 
The increase
in the temperature of the
cosmic microwave background radiation (CMB) at high redshift does mean that there will
be some increase in the dust temperature \citep{dacunha2013},
but the lack of evolution claimed in the studies above generally refers to the temperature
the dust would be at in the absence of the CMB e.g. \citet{ward2024}.

There is a well-known bias that when a single modified blackbody is fitted
to a set of far-infrared/submillimete photometry for a galaxy, the measured temperature - a luminosity-weighted temperature -
is always higher than the mean mass-weighted dust temperature \citep{eales1989,liang2019}, which is what one needs to know for an accurate
estimate of the dust mass.
The reason for this is that there is always a mixture of dust at different temperatures
in a galaxy, and since warm dust radiates more strongly, and at a shorter wavelength,
than the same mass of cold dust, the measured, luminosity-weighted, temperature is always biased high.
For a given set of observation wavelengths, this bias increases with redshift, since
as the redshift increases, the part of the galaxy's rest-frame spectral energy distribution
(SED) covered by the observation set will move to shorter wavelengths. The 
problem is not as bad for an estimate of the bolometric luminosity of the
dust, since the measured, luminosity-weighted, temperature is much closer to what one needs for an accurate estimate of the
bolometric luminosity.

In Appendix B, we show that this same bias applies to temperature measurements
from stacking analyses, which sometimes show dramatic increases in dust temperature
with redshift \citep{viero2022}. We show that this bias
is present even if all the individual galaxies in the stacking sample only
contain dust at a single temperature. For this reason, the temperature models below are
based on observations of individual galaxies rather than estimates in the literature of the
dust temperature from stacking analyses.

We have explored the sensitivity of our estimates of dust mass and bolometric dust luminosity
using three temperature models:

\begin{itemize}

\item {\bf COLD:} In this model, we assume that the mean mass-weighted dust temperature is 22 K and
does not change with redshift, which is the assumption we made in
our previous paper \citep{eales2024b}. This is the median dust temperature found from the
fits of a single modified black body to the flux densities for $\sim$80,000 galaxies
found in the Herschel ATLAS survey \citep{eales2024b}. There was no evidence in this study
for any variation of dust temperature with redshift out to $\rm z \simeq 1$.
Despite the disagreement between the observational studies, cosmological
simulations have show that at higher redshifts, while the observed, luminosity-weighted, dust temperature does increase with redshift, the
mass-weighted dust temperature does not \citep{liang2019}. We therefore assume in this model that there is no evolution
in the dust temperature.

\item {\bf STAR-FORMER:} The first model must be, at least, a simplification of the
situation in real galaxies, because unless a galaxy contains no newly formed stars
and no active nucleus, it must contain some phase of warmer dust.
The galaxies that seem likely to contain the highest fraction of warm dust 
are the luminous dusty star-forming galaxies, which are often
labelled as ultraluminous infrared galaxies at low redshift 
and SMGs at high redshift. 
Models of the spectral energy distributions (henceforth SEDs) that include two modified black
bodies have shown that although the galaxies in these classes do contain
some warm dust, most of the dust is
at a low temperature \citep{dunne2001,pearson2013,bakx2018}.
As our second model,
we have adopted the model of \citet{bakx2018}, who found that the average SED of the most
luminous SMGs at $\rm z>2$ could be represented by a mixture of dust at 21.3 K and
45.8 K, with the ratio of the mass of cold dust to warm dust being 26.6.
Since this is very similar to the results for low-redshift ultraluminous infrared galaxies
\citep{dunne2001}, we include no evolution in this model.

\item {\bf EVOLVE:} This model is based on the dust temperatures measured for 17 galaxies
over the redshift range $\rm 4 < z < 8$ \citep{witstok2023}, which do show an
increase of the dust temperature with increasing redshift. The relationship matches well the prediction
from cosmological simulations for the relationship between 
the observed, luminosity-weighted, dust temperature and redshift \citep{liang2019}. We have used the
relationship predicted from the cosmological simulations as our model, with the
small modification that we have used a temperature of 23 K on the left-hand side of
the relationship, rather than the 25 K found in the simulations \citep{liang2019},
in order to make it consistent with our first two models at low redshift:

\begin{equation}
log_{10}\left( {T_{dust} \over 23\ K}\right) = -0.02 + 0.25 log_{10}(1+z)
\end{equation}

We note that although this relationship shows evolution and matches well
the observations of the high-redshift sample \citep{witstok2023}, the same
simulations predict that although the measured, luminosity-weighted, dust temperature does
change with redshift, the mass-weighted dust temperature, which
is what we really want, does not change
with redshift \citep{liang2019}.
\end{itemize}

\begin{figure*}
	\includegraphics[width=\textwidth]{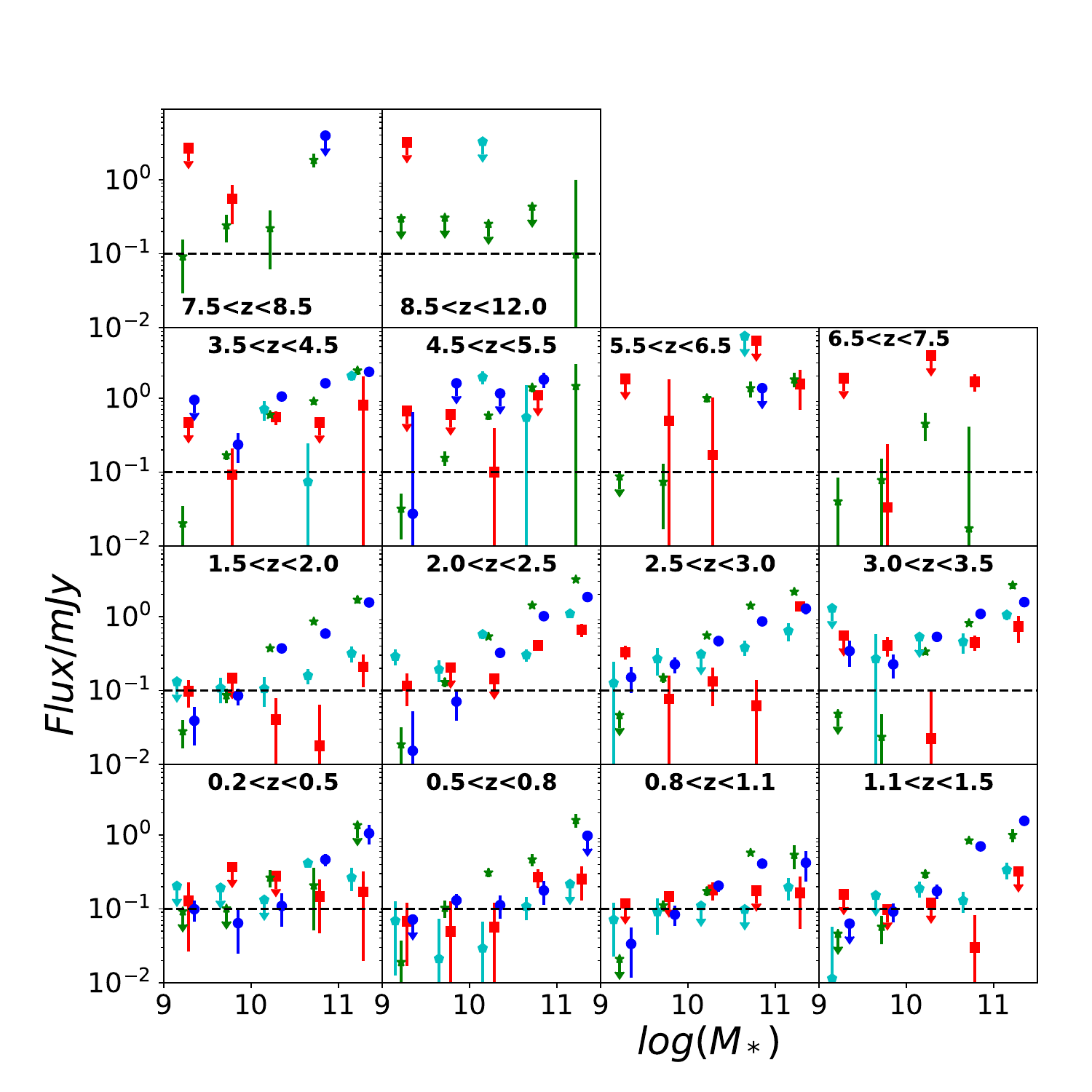}
	\caption{Mean flux density at 850 $\mu$m estimated with SIMSTACK for the four different
    morphological classes from \citet{huertas2025}: irregular galaxies--green; disk-dominated galaxies--blue; bulge-dominated
    galaxies--cyan; spheroids--red. The horizontal dashed line at 0.1 mJy has
    been drawn to aid comparison between the panels. We have added
    offsets to the sets of points
    (0.05 in $\rm log_{10}(M_*)$) to avoid them overlapping.
    The upper limits are 3$\sigma$ upper limits.}
	\label{fig:morphologies}
\end{figure*}

We believe the model that is likely to be closest
to the truth is {\bf STAR-FORMER}, since 
it is based on a more sophisticated model than a single modified blackboday, 
making it less susceptible to the temperature bias described above, and is an empirical fit to the SEDs of real star-forming galaxies. 
We have included the other two models to show the effect on the dust masses
of the range of dust temperatures claimed in the literature. 

The models above describe the temperature of the dust in the absence of
the effect of the cosmic microwave background (CMB). We combined the temperature model
and the effect of the CMB as follows. At each redshift we used the model to estimate the
temperature the dust would have in the absence of the CMB. We used equations 12 and 18
in \citet{dacunha2013} to calculate the actual temperature of the dust and
to correct the observed flux
density for its suppression in the presence of the CMB. In the case of {\bf STAR-FORMER}, we corrected the temperature and flux density for each component separately and then added them to produce the
final flux. In calculating dust masses for this model (\S 3.3), we
calculated the dust mass for each component separately and then
added them to produce the final dust mass.
The corrections can be very large. For {\bf COLD}, which is the most extreme case,
the correction factors for the flux
density are 1.04, 1.91 and 16.8 at $\rm z=3$, $\rm z=7$ and $\rm z=12$,
respectively.

Finally, in this section, we discuss the effects of using these different
models on our estimates of dust mass and bolometric dust emission, which we use
in \S4 to estimate the star-formation rate. {\bf COLD} and {\bf STAR-FORMER} both yield
very similar estimates of the dust mass because in the former all the dust and in the latter 96\% of the dust mass is at $T\simeq20\ K$. The two models, though, yield very different estimates of the bolometric dust luminosity because the small percentage of
warm dust in {\bf STAR-FORMER} vastly out-radiates the 96\% of the dust that is cold.
The estimates of the dust mass from {\bf EVOLVE} are similar to the estimates
from the other two models at low redshift where the dust temperatures are similar but get increasingly lower than the other two as the redshift, and therefore the temperature of the dust, increases. The estimates of bolometric dust emission for {\bf EVOLVE}
and {\bf STAR-FORMER} are similar at all redshifts because in the case of {\bf STAR-FORMER} it is the small percentage of warm dust that is responsible for most
of the bolometric dust emission.

\section{Results} \label{sec:results}

\subsection{Comparison between the results of the two stacking methods}

Figure 1 shows the signal-to-noise images for each sample of galaxies and for each stacking method. Figure 2
shows mean flux density versus mass for each redshift interval, 
with the results for the basic stacking
method shown in mauve and for SIMSTACK in cyan. The values of the flux density and errors obtained from SIMSTACK 
are given
in Appendix A.

Both figures show fairly good agreement between the results from the two stacking methods. Both methods
show that in all redshift intervals there is a increase in flux density with stellar mass, which is itself
a useful sense check since we can not think of any reason why a stacking method would
produce a spurious relationship like this and it
is what one expects if all galaxies have roughly the same ratio of dust mass to stellar mass.

The differences between the results from the two methods are instructive. First, the SIMSTACK flux densities are
almost always slightly lower than those from the basic method, which
is what one would expect if
galaxy clustering is enhancing the signal measured with the latter method.
Second, while Figure 1 shows that the basic method produces
significant detections at $\rm log_{10}M_* < 10.0$, these mostly disappear in the SIMSTACK results. This is also what one would expect. Suppose there are low-mass galaxies clustered around high-mass galaxies
within a SCUBA-2 beam. The mean flux densities of each class
will be enhanced but the flux densities of the low-mass galaxies will be proportionately more enhanced.
Since the effect of
galaxy clustering has been corrected in the SIMSTACK results, these are the ones we use in the rest of the
paper.

\begin{figure*}
	\includegraphics[width=\textwidth]{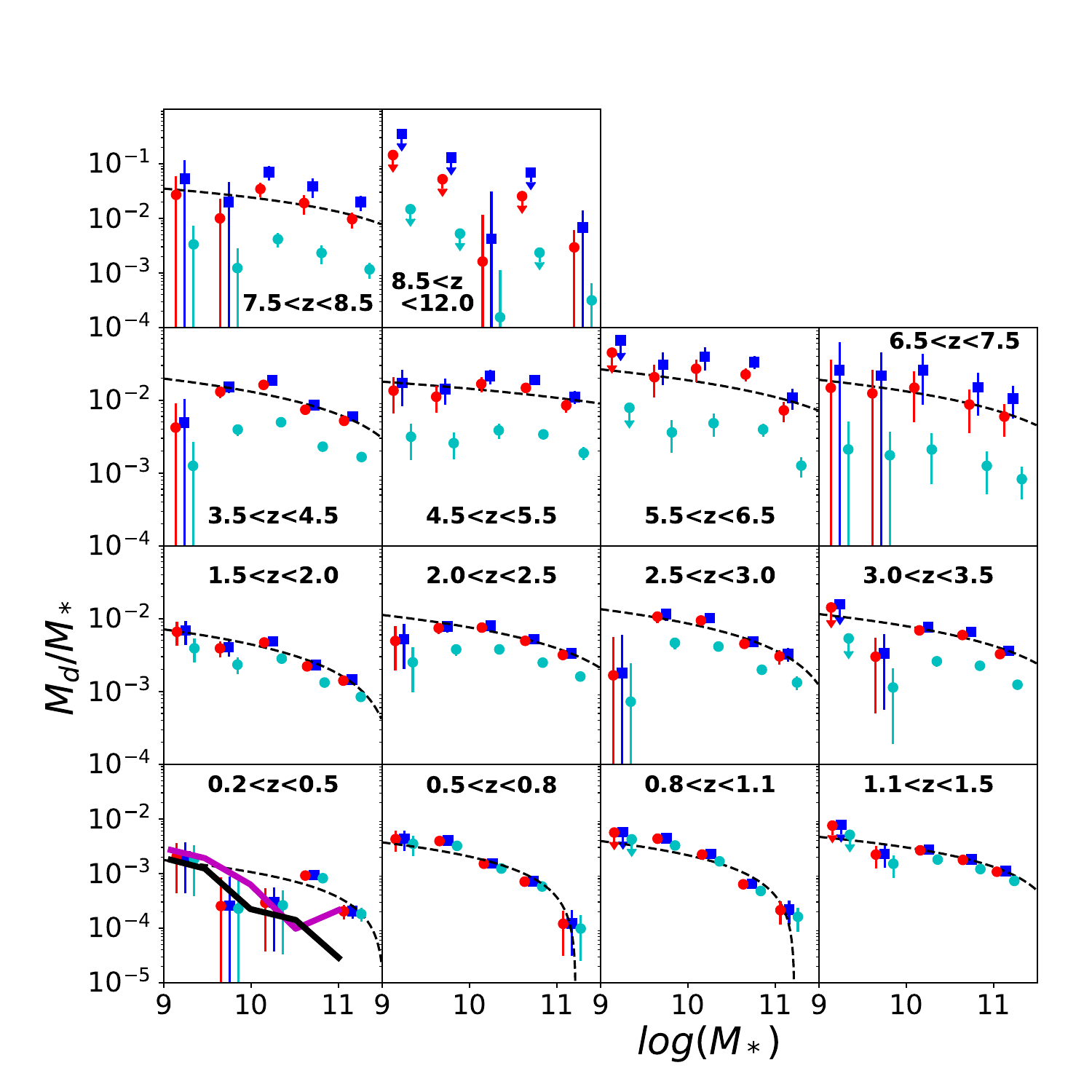}
	\caption{The mean ratio of dust mass to stellar mass versus stellar mass for the 14 redshift
    intervals. The coloured symbols show the results for the three temperature models: blue--{\bf COLD}; 
    red--{\bf STAR-FORMER}; cyan--{\bf EVOLVE}. The three sets of points have been offset
    by $\pm$0.1 along the x-axis for greater clarity.The upper limits are 3$\sigma$ upper limits. The dashed lines
    show the result of fitting equation 5 to the data points for
    {\bf STAR-FORMER}.
    The two thick lines in the lowest redshift interval shows relationships
    derived for the Herschel Reference Survey \citep{cortese2012} after
    a correction has been made for the different value of the mass-opacity
    coefficient. The purple line is for galaxies outside the Virgo Cluster,
    the black line for galaxies inside the Virgo Cluster.}
	\label{fig:dust density}
\end{figure*}

\subsection{Comparison with previous results}

There have been two previous stacking analyses of the S2COSMOS images, which both
used SIMSTACK but with galaxy catalogues from ground-based telescopes. 
Our results are very similar
to those of \citet{simpson2019}. We cannot compare our results precisely with the results 
of the other study \citep{Millard2020} because of the different binning procedure and because the earlier
paper did not include the errors on the SIMSTACK fluxes. Although the same trends of flux density with
stellar mass and redshift are seen in the two studies, our measurements are lower by $\sim30\%$ than 
the fluxes in the earlier paper. We do not know the explanation of the difference.

We have made an additional check on our results by comparing them to the results of 
stacking analyses of 
massive galaxies with very low star-formation rates, `massive quiescent galaxies'. We divided our samples into star-forming 
and quiescent galaxies using the LePHARE estimates (\S 2.1) of the
specific star-formation rate
(star-formation rate divided by stellar mass - SSFR).
Star-forming galaxies mostly lie on a strip
across a diagram of SSFR versus stellar mass, which is often called
the `galaxy main sequence'. We have used a relationship for the
evolution of the galaxy main sequence \citep{speagle2014} to divide the galaxies into the two classes,
classifying a galaxy as quiescent if its SSFR is at least 1.5 orders of magnitude below 
a galaxy on the galaxy main sequence with the same stellar mass.
We applied the two stacking methods (\S2.5 and 2.6) to
the two classes, each split into the same 70 bins of stellar mass and redshift
as before. In the case of SIMSTACK, we applied the algorithm to all 140 samples
simultaneously. We again carried out a Monte-Carlo simulation to estimate the errors in the SIMSTACK
flux densities
arising from the errors in the COSMOS-Web photometric redshifts and stellar mass estimates,
adding these errors in quadrature to the basic SIMSTACK errors.

Figure 3 shows the results. The star-formation rate and the mass of dust in
a galaxy are likely to be correlated because both are correlated with the
mass of gas in the galaxy \citep{kennicutt1998,eales2012}. One might therefore
expect the submillimetre flux density of the star-forming galaxies to be higher
than for the quiescent galaxies in the same redshift and mass interval, and this is what one generally but not always sees in the panels.

In Table 2 we compare our results with other attempts to estimate, through stacking, the
mean 850-$\mu$m flux density of samples of massive ($\rm M > 10^{11}\ M_{\odot}$) quiescent galaxies 
\citep{gobat2018,magdis2021}. Our estimates are consistent with the previous estimates.

\begin{table}
	\centering
	\caption{Comparison with previous results for quiescent galaxies}
	\label{tab:example_table}
	\begin{tabular}{ccl} 
		\hline
		Redshift range & Mean 850-$\mu$m flux density & Paper\\
		\hline
		$\rm z \sim 1.76$ & 81$\pm$23 $\mu$Jy & \citet{gobat2018}\\
        $1.5 < z < 2.0$ & 147$\pm$102 $\mu$Jy & This paper \\
        $\rm <z>=0.5$ & 100$\pm$40 $\mu$Jy & \citet{magdis2021} \\
        $\rm 0.2 < z < 0.5$ & 156$\pm$131 $\mu$Jy & This paper \\
        $\rm <z>=0.85$ & $<$210 $\mu$Jy & \citet{magdis2021} \\
        $\rm 0.5 < z < 0.8$ & 109$\pm$102 $\mu$Jy & This paper \\
        $\rm <z>=1.20$ & 40$\pm$30 $\mu$Jy & \citet{magdis2021} \\
        $\rm 1.1 < z < 1.5$ & 55 $\pm$105 $\mu$Jy & This paper \\
		\hline
	\end{tabular}
\end{table}

\begin{figure*}
	\includegraphics[width=\textwidth]{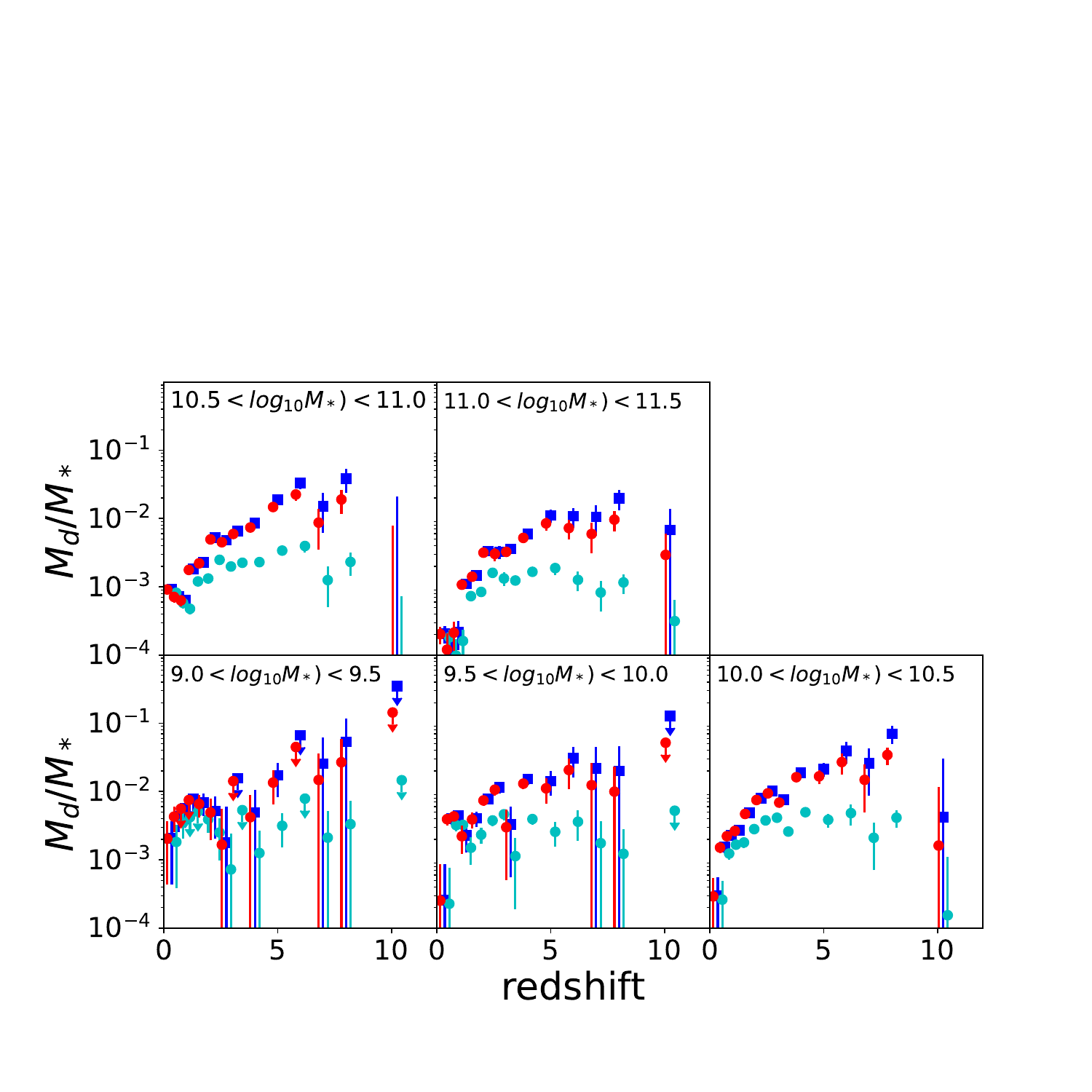}
	\caption{The mean ratio of dust mass to stellar mass versus redshift for the five intervals
    of stellar mass. The coloured symbols are for the different temperature models: blue--{\bf COLD};
    red--{\bf STAR-FORMER}; cyan--{\bf EVOLVE}. The sets of points have been offset by $\pm$0.1 in redshift for
    greater clarity.}
	\label{fig:dust density}
\end{figure*}

\subsection{The results for different morphological types}

We applied  the two stacking methods to the four morphological classes: spheroidal, disk-dominated, irregular
and bulge-dominated. We split each class into the same 70 bins
of redshift and stellar mass as before. In the case of SIMSTACK, we applied
the algorithm to all 280 samples simultaneously. We again carried out a Monte-Carlo simulation to estimate 
the errors in the SIMSTACK results
arising from the errors in the COSMOS-Web photometric redshifts and stellar mass estimates,
adding these errors in quadrature to the basic SIMSTACK errors.

Figure 4 shows the results.
Today, spheroidal and bulge-dominated galaxies
contain less dust than disk-dominated and peculiar galaxies, with spheroidal galaxies, in 
particular, containing very little dust \citep{smith2012}.
The figure shows that in most panels the submillimetre emission from disk-dominated and
irregular galaxies is generally higher than for bulge-dominated and spheroidal
galaxies, but the emission from all the morphological categories increases
with redshift in roughly the same way. At $\rm z > 4$, 
the regular Hubble sequence has largely vanished \citep{huertas2025}, and 
the only
samples that contain enough galaxies for a measurement are the irregular
galaxies.

\subsection{The ratio of dust mass to stellar mass}

We calculated the mean ratio of dust mass to stellar mass for each of the 70 samples of galaxies
in the following way. Starting from the mean flux density for that sample from the SIMSTACK
analysis, we estimated the dust mass for each of the galaxies in that sample using the
following equation:

\begin{equation}
    M_d = \frac{<S_{850\mu m}> D_L^2}{(1+z) B(\nu_e) \kappa(\nu_e)} 
	\label{eq:md}
\end{equation}

\noindent in which $\nu_e$ is the frequency in the rest frame: $\nu_e = (1+z) \nu_{850\mu m}$. $D_L$ is luminosity distance, and the
dust mass-opacity coefficient, $\kappa(\nu)$, is given by:

\begin{equation}
    \kappa(\nu) = \left(\frac{\nu}{\nu_{850 \mu m}} \right)^{\beta} \kappa_{850 \mu m} 
	\label{eq:kappa}
\end{equation}

\noindent We assumed a value for $\beta$ of 2, which is the value found in many
recent studies of high-redshift galaxies \citep{dacunha2021,ismail2023,bendo2023,witstok2023,ward2024,bendo2025}, and 
the widely used value for the mass-opacity coefficient of
$\kappa_{850\mu m} = 0.077\ m^2\ kg^{-1}$ \citep{james2002,dacunha2008,dunne2011}. The value of the mass-opacity coefficient is notoriously uncertain
\citep{clark2016}, but we note that our value was calibrated directly
from estimates of the mass of metals in a galaxy \citep{james2002}. This makes
it an appropriate one to use if one 
wants to use a model of the evolution of the metals in a galaxy
to also model the evolution of the dust, which we do in \S 4.5.

Before calculating the dust mass, we corrected the flux density and the
dust temperature for the effect of the cosmic microwave background (\S2.7).
We calculated the dust mass for all three temperature models: {\bf COLD}, {\bf STAR-FORMER} and
{\bf EVOLVE} (\S2.7).

Once we had calculated a dust mass for each galaxy in the sample, we calculated the mean value of the ratio of the
dust mass to stellar mass for all the galaxies in the sample, using the values of the stellar mass for each
galaxy given in the COSMOS-Web LePHARE catalogue (\S 2.2).

Figure 5 shows the mean ratio of dust mass to stellar mass versus stellar mass for each of the 14 redshift
intervals. The estimates for {\bf COLD} and {\bf STAR-FORMER} are fairly similar
at all redshifts, but the estimates for {\bf EVOLVE} are lower than for the other two, with the difference
increasing with redshift. For all three models, at low redshifts ($\rm z \leq 3$) the ratio of dust mass to stellar mass
decreases with increasing stellar mass. At higher redshifts, the mass ratio is roughly independent of stellar mass.
For each redshift interval, we fitted the
following relationship to the data points for {\bf STAR-FORMER}:

\begin{equation}
    \frac{M_*}{M_s} = a log_{10}M_* + b
	\label{eq:fit 1}
\end{equation}

\noindent The fits are shown in the figure and Table 1 lists our estimates of $a$ and $b$.

Figure 6 shows the mean ratio of dust mass to stellar mass plotted against
redshift for each interval of stellar mass. 
For all three temperature models, there is a
steady increase of the mass ratio with redshift in all mass intervals
except in the two lowest mass intervals for {\bf EVOLVE}.

Figure 7 shows the weighted mean value of the 
five values for the dust-to-stellar mass ratio in each redshift interval plotted
against redshift. We have calculated these weighted means separately for each
temperature model and used the errors on dust-to-stellar mass ratios shown in
Figure 6.
For all three temperature models, the mass ratio increases with redshift out to $\rm z \sim 2.5$.
For {\bf COLD} and {\bf STAR-FORMER} the mass ratio continues to increase with redshift at $\rm z > 2$
but is fairly constant above this redshift for {\bf EVOLVE}.

\begin{figure}
	\includegraphics[width=\columnwidth]{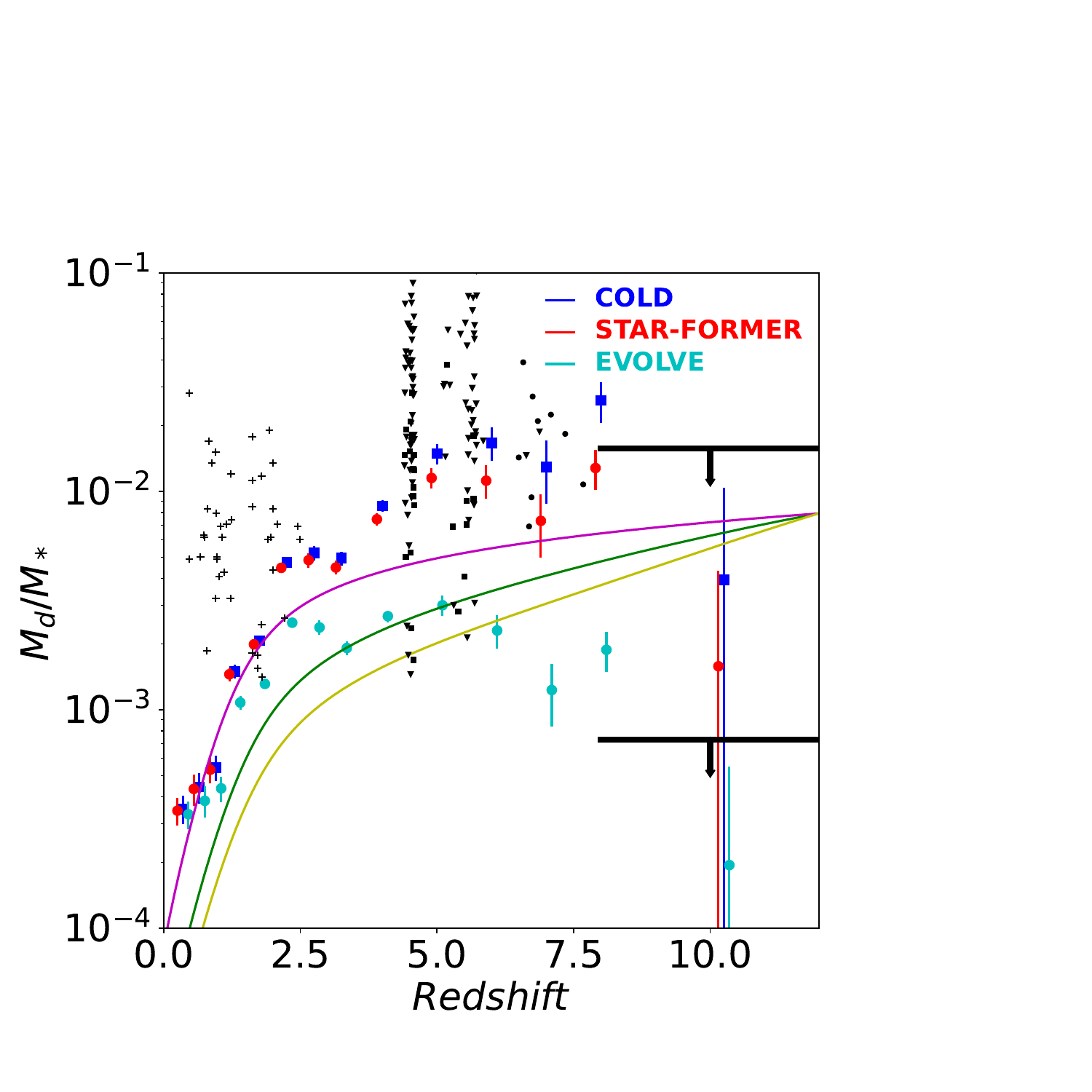}
	\caption{The mean ratio of dust mass to stellar mass (see text for details)
    plotted against redshift. The coloured symbols are for the
different temperature models: blue--{\bf COLD}; red--{\bf STAR-FORMER}; cyan--{\bf EVOLVE}. The black symbols show estimates of the dust-to-stellar mass ratio for individual galaxies: crosses - \citet{kirkpatrick2017}; squares - \citet{bethermin2020}; squares - \citet{algera2026} (see text for more details).
The lower horizontal line
shows the upper limit for the mass ratio derived by \citet{bakx2026}
for galaxies at $z > 8$ and the upper limit shows our
recalculation of this limit if we assume a dust temperature of 22 K.
    The three coloured lines are the predictions of our model for the evolution of the
mass ratio (\S4.5), with the purple line showing the prediction for a closed-box model ($\Lambda = 0$),
 the green line for a model with outflows and $\Lambda = 1$, and the yellow line
 for a model with outflows and $\Lambda=2$.}
	\label{fig:dust density}
\end{figure}

We have compared our results at low redshift with the dust-scaling relations derived
for the Herschel Reference Survey (HRS) \citep{cortese2012}. These authors used a method
that is sensitive to the cold dust in a galaxy, which should produce similar
results to our {\bf COLD} and {\bf STAR-FORMER} temperature models, and
also used the same value for the dust emissivity index as ours. They used a different mass-opacity coefficient, but we have made a correction to
their relationships to account for this. In the panel for the lowest redshift interval in Fig. 5 we have shown their relationships between the ratio of dust mass to stellar mass and stellar mass for the HRS galaxies inside and outside the Virgo Cluster. Apart from our measurement for the
stellar mass bin with $10.5 < log_{10} M_{*} < 11.0$, the HRS relationships agree
very well with ours. The HRS relationships show the same decrease of the mass ratio
with stellar mass that we find at all redshifts at $z < 3.5$

In Figure 7 have compared our estimates of the mean ratio of
dust-to-stellar mass with estimates 
for individual galaxies from three programmes. At $z>6$ we have used estimates
of the ratio of dust mass to stellar mass in \citet{algera2026}, which are based on
ALMA continuum measurements of galaxies in the Reionization Era Bright Emission Line Survey (REBELS; \citet{bouwens2022}). These estimates were made on the assumption of a dust temperature of 45K \citep{algera2026}. We have scaled these estimates to the values they would have if they had been based on our {\bf STAR-FORMER} temperature model
and our value of the mass-opacity coefficient. 
At slightly lower
redshifts ($4.4<z<5.9$), we have used the ALMA continuum measurements of galaxies
in the ALMA Large Program to Investigate [CII] at Early times (ALPINE, \citet{lefevre2020,bethermin2020}). We have used these measurements \citep{bethermin2020} plus the
redshifts \citep{faisst2020} to estimate dust masses using our value for
the mass-opacity coefficient and the {\bf STAR-FORMER} temperature model, calculating
the ratio of dust mass to stellar mass using the stellar
mass estimates in \citet{faisst2020}. In the redshift range $0<z<2$ we have used
estimates of the ratio of dust-to-stellar mass for several samples of galaxies with high far-infrared luminosities in \citet{kirkpatrick2017}, who used {\it Herschel} and {\it Spitzer}
measurements to estimate the dust masses. We have converted these to the same
value of the mass-opacity coefficient that we use in this paper\footnote{\citet{kirkpatrick2017} estimated their dust masses using the
model in \citet{weingartner2001}, in which the mass-opacity coefficient
depends on wavelength in a complex way. We have used the approximation
for this relationship given in \citet{liang2019}.}.

Figure 7 shows that the estimates from REBELS of the dust-to-stellar mass ratio
are mostly higher than our estimate
for the mean value
at $6<z<8$. The estimates from ALPINE at $4.4<z<5.9$ are mostly upper limits but the detections are spread roughly equally above and below our estimate of the mean value.
The estimates of the dust-to-stellar mass ratio for the low-redshift ($0<z<2$) sample
are
consistently higher than our estimate of the mean value.
This last difference is perhaps not surprising because these galaxies
are from samples of galaxies with high far-infrared luminosities and so there
is likely to be a selection effect favouring high values of the dust-to-stellar
mass ratio \citep{eales96}. There is not the same strong bias operating
at higher redshifts, but both REBELS and ALPINE were targeted at galaxies with
star-formation rates above a minimum value \citep{Bouwens2020,lefevre2020}, so
there may be some bias in these samples
towards higher values of the dust-to-stellar mass ratio, compared to our estimates
which are for the mean value for all galaxies within a certain range of stellar mass.

The main feature of Figure 7 is the steady rise in the mean dust-to-stellar
mass ratio with redshift. There is currently no submillimetre survey of a
sufficiently large area of sky and with sufficient sensitivity to produce
dust-mass estimates for a large percentage of the
individual galaxies in a large unbiased
sample in the redshift range $0<z<4$. However, studies like ours that use a statistical method to estimate the mean ratio
do find an increase with redshift similar to the one we see
\citep{calura2017,scoville2017,Millard2020}.

At the highest redshifts ($z>8$), \citet{bakx2026} estimated an upper limit for the mean ratio of dust mass to stellar mass. We have plotted their upper limit (for galaxies with $\rm log_{10} M_* > 9.0$) in Figure 7, after making a correction
for the different value of the mass-opacity coefficient, as the lower horizontal
line in the figure. At first sight, this disagrees with our estimates in the
redshift bin $7.5 < z < 8.5$. However, in estimating this limit, \citet{bakx2026} assumed a dust temperature of 50 K. The upper horizontal line
shows their upper limit if we recompute it using {\bf COLD} ($T = 22\ K$). The
dramatic difference shows the extreme sensitivity of dust-mass estimates at high redshift to the temperature model.

\subsection{The evolution of the cosmic dust density}

As in our previous paper \citep{eales2024b}, we used the stacking measurements and the stellar mass function
to calculate the mean density of dust in the universe in each redshift interval.
We calculated the contribution to the dust density from the i'th sample, $\rho_i$, from
the equation:

\begin{equation}
    \rho_i = \frac{< {M_d \over M_*}> \int_{z_L}^{z_U} \int_{M_L}^{M_U} \phi(M_*) M_* dM_* dV}{\int_{z_L}^{z_U} dV}
	\label{eq:dust density}
\end{equation}

\noindent in which $\rm M_*$ is the mass of stars in a galaxy and the integrals are over the mass and redshift
limits for that sample. The mean ratio of dust mass to stellar mass that precedes the integral is our estimate for
that sample (\S3.4). We carried out the integration over the mass
range $9.0 < log_{10}M_* < 11.5$. We made no attempt to correct
for the galaxies with masses outside this range because the corrections are
so uncertain. In our previous paper \citep{eales2024a}, we did make an
attempt to do this, finding corrections of $\simeq10-20\%$ to the mean
dust density.

In applying equation 6, we used the tabulated values of the stellar mass function derived from
the COSMOS-Web data \citep{shuntov2025b}. The redshift ranges we have used for
the samples (Table 1) are the same as those used to derive the COSMOS-Web stellar mass function,
except that to maximise signal-to-noise at high redshifts we combined the two highest redshift
intervals used to derive the stellar mass function into a single redshift interval.
To derive the mean dust density in this redshift interval, we have therefore
averaged the two
COSMOS-Web stellar mass functions.

The mean density of dust in each redshift interval is
then given by:

\begin{equation}
    \rho = \sum_i \rho_i 
	\label{eq:dust density 2}
\end{equation}

\noindent in which the sum is over all the stellar-mass bins for that redshift interval.

Figure 8 shows the results. There is good agreement between the mean dust density derived
with {\bf COLD} and {\bf STAR-FORMER}, 
which is not surprising because most of the dust in both models is cold. The estimates
for
{\bf EVOLVE} are lower, with the difference increasing with redshift.  If
the temperature estimates made from some of the stacking analyses turn out to be correct,
the differences would be even larger.
For example, if we used a temperature of 100 K for dust at $\rm z \sim 7$
\citep{viero2022}, our estimate of the global dust density at this redshift
would be an additional factor of $\sim17$ lower 
than the lowest of our three estimates.

The purple circles
show the
estimates from our earlier study, which used {\bf COLD} as the temperature model \citep{eales2024b}.
There is good overall agreement between our new estimates based on COSMOS-Web and our old
estimates based on COSMOS. The significant differences at some redshifts (e.g. $\rm 1.1 < z < 1.5$)
are mostly explained by differences in the stellar mass functions used in the two
studies, which are probably the result of cosmic variance.

In our earlier paper \citep{eales2024b}, we compared our estimates of the mean
dust density with all other estimates we were aware of, and
we refer the reader to that paper for a comparison of our results
with other results. In the earlier paper, we overlooked one study
\citep{magnelli2020}, which was based on ALMA observations of the Hubble Ultra Deep
Field. These estimates are shown in Figure 8, which are consistent with ours although
with much larger error bars.

\begin{figure}
	\includegraphics[width=\columnwidth]{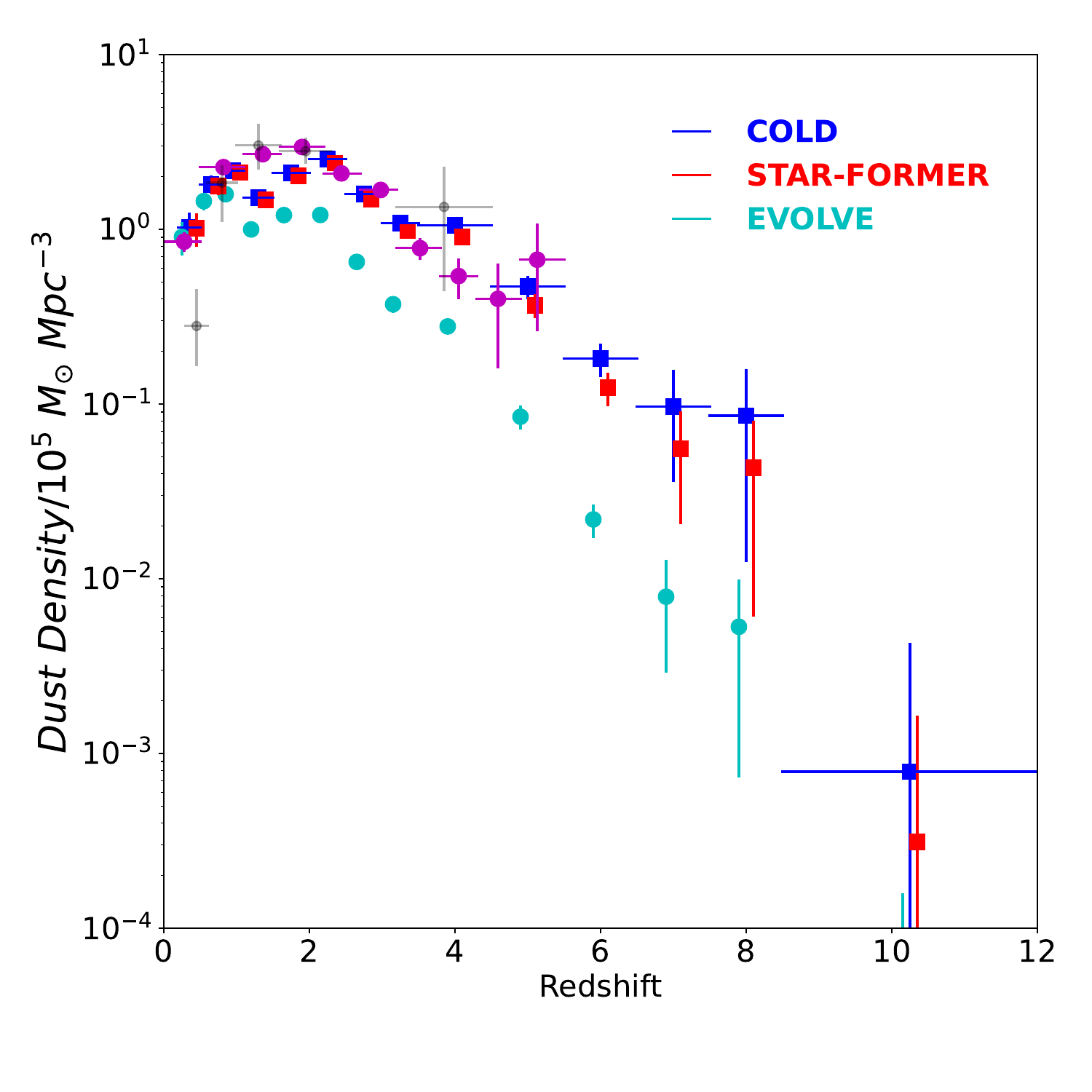}
	\caption{The mean dust density versus redshift for {\bf COLD} (blue),
    {\bf STAR-FORMER} (red) and {\bf EVOLVE} (cyan).
    The purple circles
    show our estimates based on the ground-based surveys of COSMOS \citep{eales2024b}. The
    grey crosses show the estimates of \citet{magnelli2020}.
    A comparison of our results with other attempts to derive the
    relationship between mean dust density and redshift is given in our
    earlier paper \citep{eales2024b}}.
    \label{fig:dust density}
\end{figure}

\begin{figure}
	\includegraphics[width=\columnwidth]{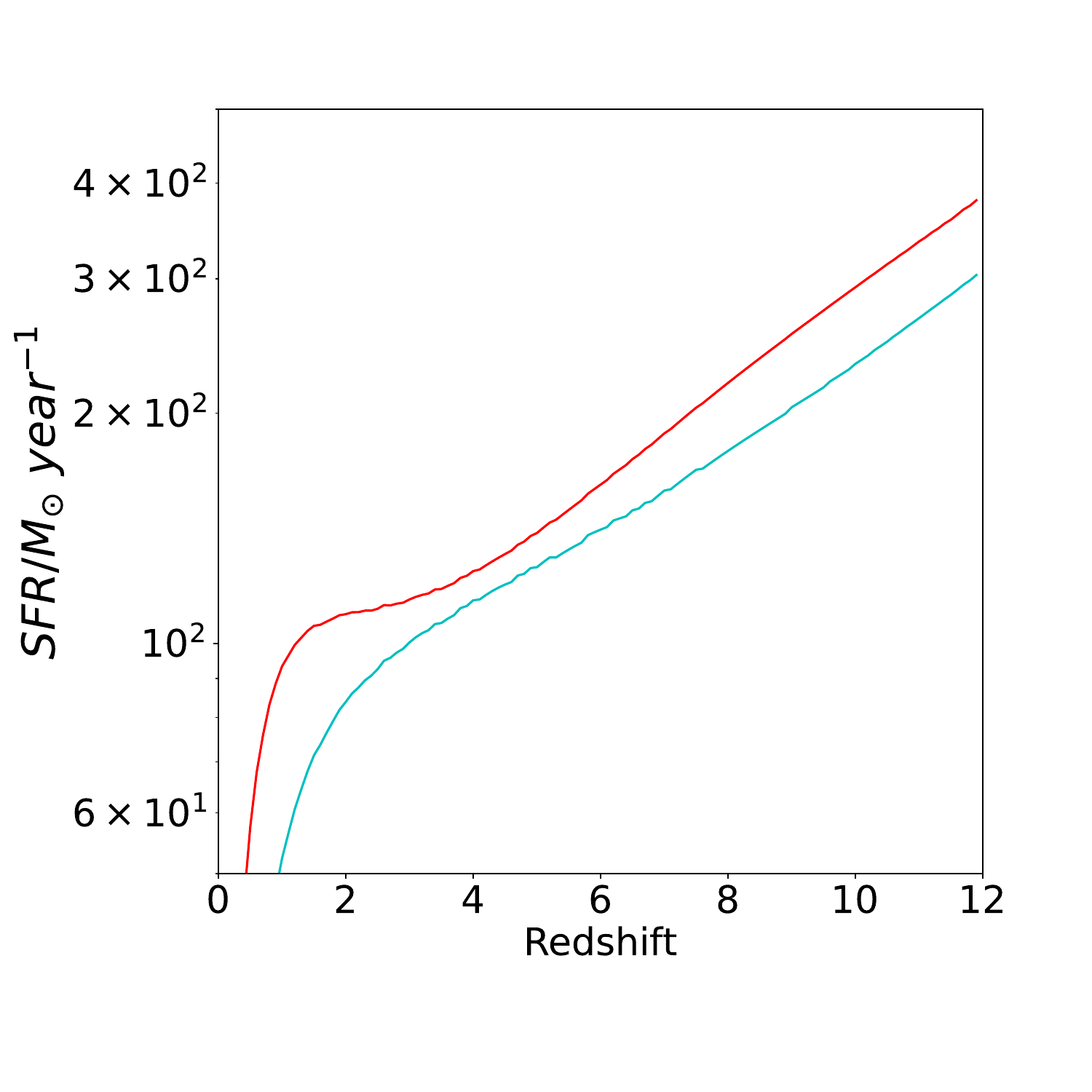}
	\caption{Estimated star-formation rate versus redshift for a galaxy with an 850-$\mu$m flux density
    of 1 mJy (see text for details) and for the temperature models: 
    {\bf EVOLVE (cyan)} and {\bf STAR-FORMER (red).}}
	\label{fig:morphologies}
\end{figure}

\section{Discussion} \label{sec:discussion}

\subsection{The star formation rate estimated two ways}

In this section, we switch our attention from dust masses to the bolometric dust
luminosity. The reason for this is that in the context of galaxy evolution two
of the most important properties of a galaxy are its star-formation rate and
its gas mass. Although the dust mass is often used to estimate the mass of gas in
a galaxy \citep{eales2012,scoville2016,scoville2017,tacconi2020}, the
method depends on an assumption about the gas-to-dust ratio, which depends
on the galaxy's metal abundance and which is a particular problem at high
redshift where in general galaxies are likely to be less chemically evolved.

\begin{figure*}
	\includegraphics[width=\textwidth]{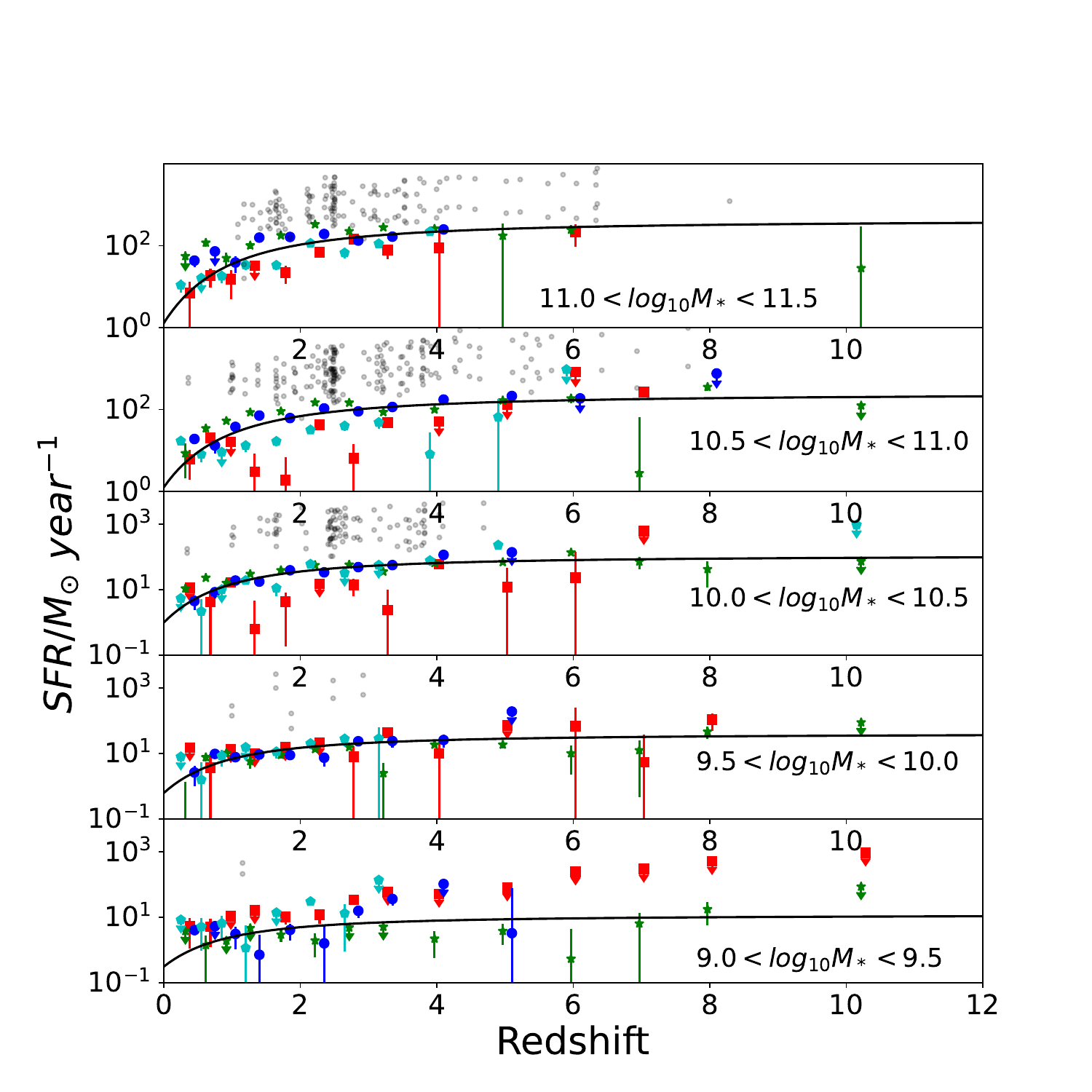}
	\caption{Mean value of the star-formation rate estimated from the bolometric
    dust luminosity plotted against redshift, with the colour of the symbol showing the morphological
    class: red--spheroids; blue--disk-dominated galaxies; green--irregular 
    galaxies; cyan--bulge-dominated galaxies. The lines show the predicted
    relationships for the `galaxy main sequence' \citep{popesso2023}.
    The upper limits are 3$\sigma$ upper limits.
    The grey points are the 
    SMGs in COSMOS-Web \citep{mckinney2025}.}
	\label{fig:morphologies}
\end{figure*}

The bolometric dust
luminosity, on the other hand, is one of the most direct ways of estimating the
star-formation rate in a galaxy \citep{kennicutt1998,kennicutt2012}. Newly formed OB massive stars spend
a large portion of their 
lives hidden by dust and so a measurement of the bolometric
dust emission provides a direct measurement of their
energy output.
In galaxies with low star-formation rates, the estimate of the star-formation rate
from the bolometric dust emission may be too high because of
the heating effect of the older stellar population
(e.g. \citet{ford2013}), but in galaxies with 
high star-formation rates - the most extreme
examples being the SMGs - this is not likely to be
a problem \citep{kennicutt1998,kennicutt2012}.

With a flux density measurement at only a single wavelength, we do need a model
of the temperature of the dust.
Figure 9
shows the star-formation
rate plotted against redshift for
a source with an 850-$\mu$m flux density of 1 mJy for 
{\bf STAR-FORMER} and {\bf EVOLVE}. We have not included
{\bf COLD} because any galaxy containing OB stars must contain some
warm dust. We estimated the star-formation rate by integrating
the SED of the dust model between wavelengths of 3 and 1100 $\mu$m
to calculate the bolometric dust emission, then calculating the star-formation
rate using the calibration in \citet{kennicutt2012}.
This calibration was based on a Kroupa initial mass function,
but a calibration based on a Chabrier initial mass function gives
a very similar result \citep{kennicutt2012}.

The two temperature models give very similar results. We chose to use
{\bf STAR-FORMER} to estimate the mean star-formation rates for
the COSMOS-Web galaxies for the reasons described in \S 2.7,
but it would have made very little difference if we had used {\bf EVOLVE}.

Figure 10 shows the mean star-formation rate plotted against redshift
for the 280 samples, split into a separate panel for each mass interval.
Many of the samples contain
so many galaxies that the formal errors are very small, making the error bars
invisible. The lines
in the plot show the predicted relationships for the `galaxy main sequence', using
the relationships from \citet{popesso2023} (equation 10 in that paper).

Figure 10 shows that the mean star-formation rate for all morphological classes
increases with redshift, roughly following the prediction for the `galaxy main sequence'. Even spheroidal galaxies, which today have very low star-formation rates, had star-formation rates that reached $\simeq100\ M_{\odot}\ yr^{-1}$ at $z>2$. In our figure, the spheroids and bulge-dominated galaxies in the lowest redshift
bin do have significant star-formation rates, but since this bin covers the redshift range $0.2<z<0.5$, it is not really a local sample, and the {\it Herschel} ATLAS
survey has shown there is strong evolution in the star-formation and
dust masses of galaxies over the redshift range $0<z<0.5$ \citep{dye2010,dunne2011,eales2018,beeston2024}.

Our conclusion that high-redshift spheroids have, on average, high star-formation
rates is
not inconsistent with the
claim that there is a large population of massive quiescent galaxies
at high redshift \citep{glazebrook2017,santini2021,carnall2023a,carnall2023b,nanayakkara2024,nanayakkara2025},
since none of these studies separated galaxies by morphological type.

We have also included in the plot estimates of the
star-formation rate for the submillimetre galaxies (SMGs) in COSMOS-Web.
The term SMG was coined to refer
to the galaxies discovered in the first submillimetre surveys, so it does
not have a clear physical definition, but a practical definition comes
from the flux limit of these original surveys: $\simeq$3 mJy \citep{hughes98,barger98,eales99}. We estimated the star-formation
rates for the SMGs in COSMOS-Web \citep{mckinney2025} using the 850-$\mu$m flux densities of the individual SMGs and the same method as for the stacked
samples. Given the difference between our stacked signals and the detection limit
for the original submillimetre surveys, it is not surprising that the SMGs have much
higher star-formation rates than our morphological samples.

We note the important point that this difference 
does not necessarily mean the SMGs are a physically distinct
class of galaxies from the general population, since our star-formation rate 
for the morphological classes are estimates of the mean star-formation rate
for each class, so the SMGs might simply be extreme members of a class
rather than something unique.
Table 3 shows the number of SMGs
in COSMOS-Web that fall in the different morphological classes. There is a much
higher fraction of irregular and disk-dominated galaxies than in the general
population, although there are still significant numbers of spheroids and
bulge-domimnated galaxies.

Figure 11 shows
the ratio of our estimate of the mean star-formation rates to the mean of the COSMOS-Web estimates
for the same galaxies, which were derived from fits of
stellar synthesis models to their UV-to-near-IR SEDs
(\S 2.2) plotted against redshift. We have also plotted this ratio for the SMGs. The figure shows
that for the SMGs the estimate from the dust emission is often much
higher than the estimate from the UV-to-near-IR SED, showing, as is well known, 
that the dust obscuration in these galaxies is often so high \citep{simopson2017,dud2020,harrington2021}
that attempts to estimate the star-formation rate by fitting
population synthesis models to the UV-to-near-IR SEDs fail catastrophically.
On the other hand, for the COSMOS-Web galaxies in general there is 
no strong evidence for any systematic difference between the
star-formation rate estimated from the dust emission and 
one estimated from fitting a population synthesis model.

\begin{table}
	\centering
	\caption{Morphological Classification of COSMOS-Web SMGs}
	\label{tab:example_table}
	\begin{tabular}{lc} 
		\hline
		Classification & No. of SMGs \\
		\hline
		unclassified & 39 \\
        spheroids & 6 \\
        disk-dominated & 63 \\
        irregular & 137 \\
        bulge-dominated & 22\\
		\hline
	\end{tabular}
\end{table}

\begin{figure*}
	\includegraphics[width=\textwidth]{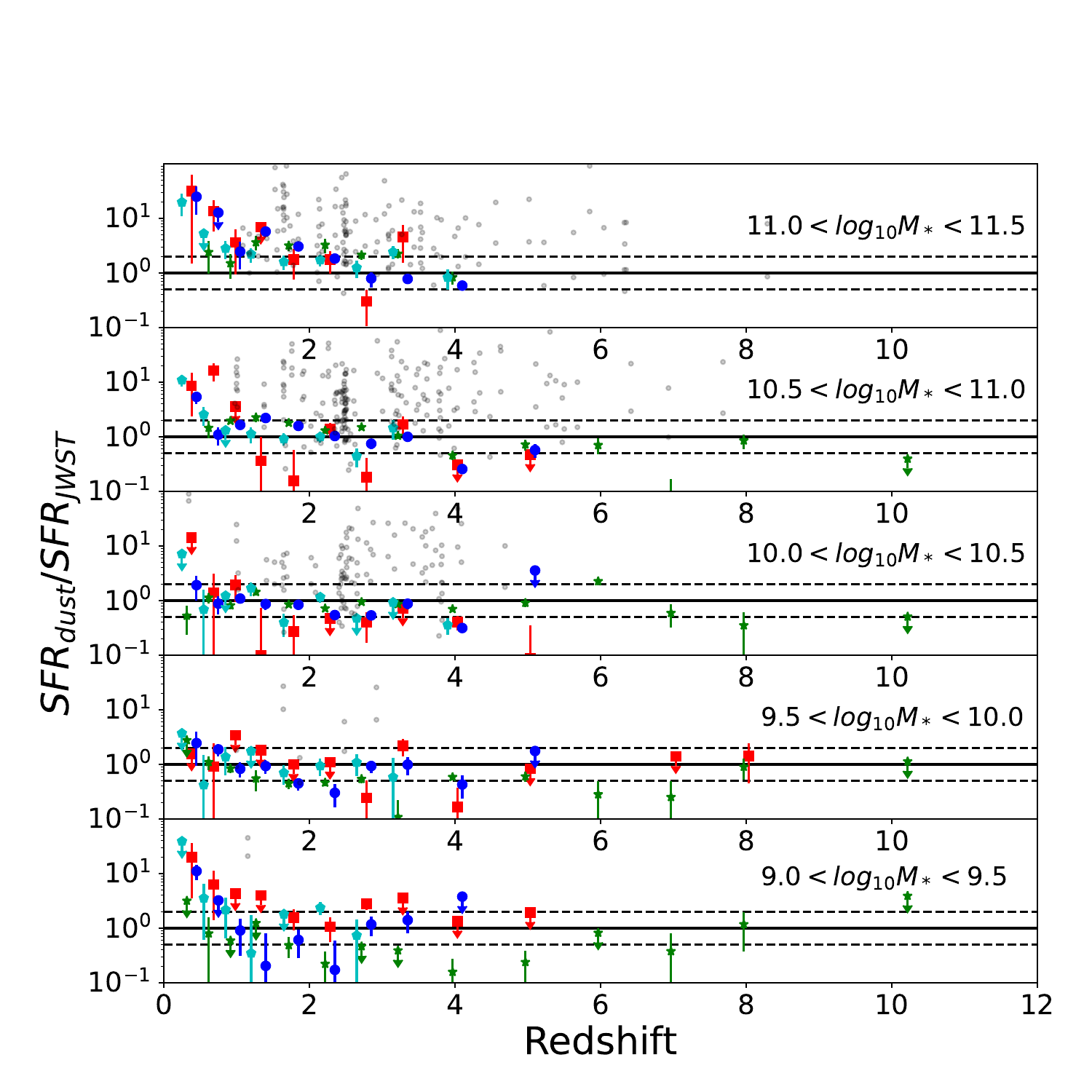}
	\caption{Ratio of the mean star-formation rate estimate from the
    dust emission to the mean of the COSMOS-Web estimates from the UV-to-near-IR SED for the same sample of galaxies.
     The continuous horizontal line shows where the two estimates
    are equal and the two dashed lines show where the star-formation rate estimated from the dust emission is
    2 or 0.5 times the COSMOS-Web estimate.
    The colours show the different morphological types: red--spheroidal
    galaxies; blue--disk-dominated galaxies; green--irregular
    galaxies; cyan--bulge-dominated galaxies. 
    The upper limits are 3$\sigma$ upper limits. 
    The grey points are SMGs in COSMOS-Web \citep{mckinney2025}.}
	\label{fig:morphologies}
\end{figure*}

\begin{figure}
	\includegraphics[width=\columnwidth]{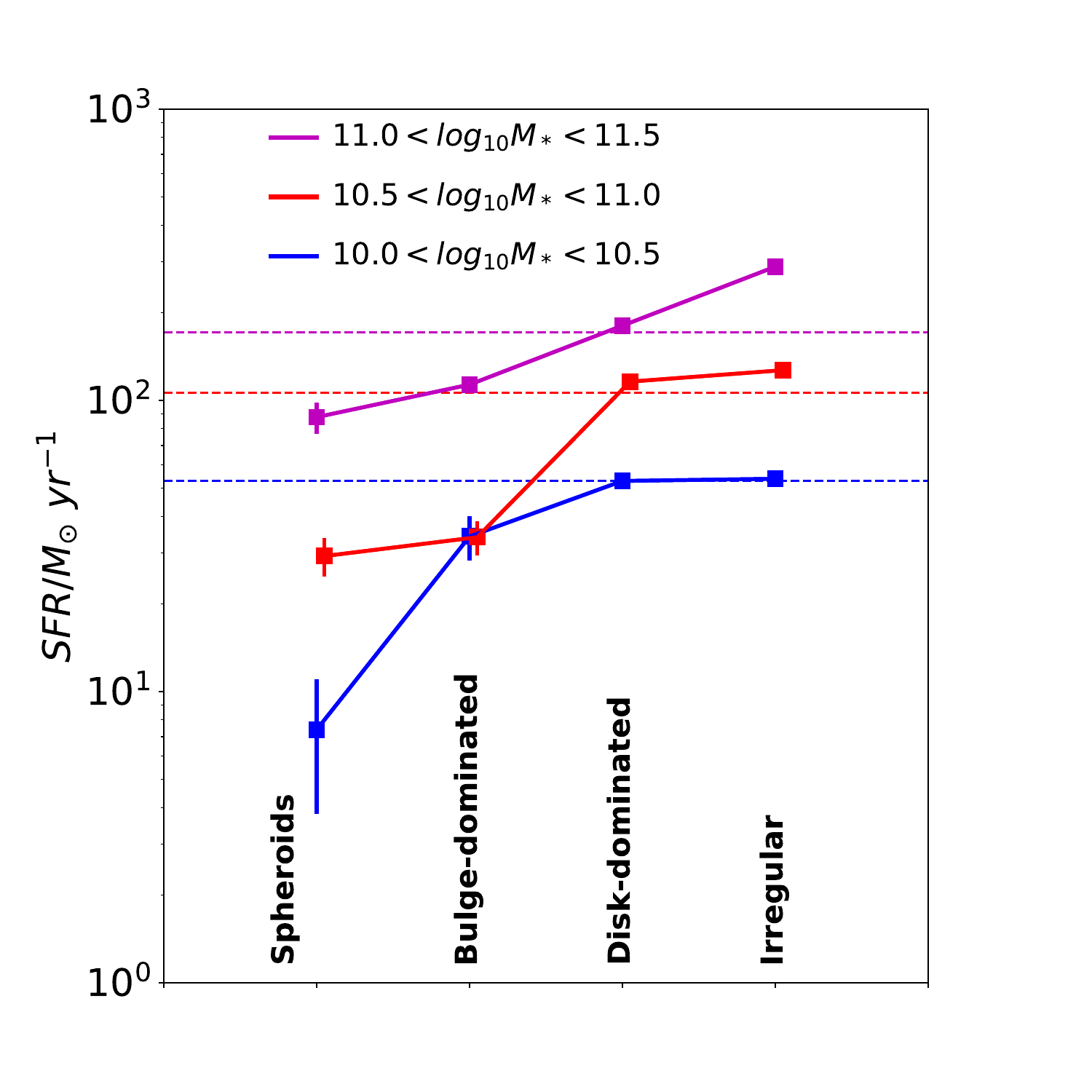}
	\caption{{Mean star-formation rate estimated from the dust emission
    for galaxies in the four morphological classs in the redshift range $2<z<4.5$ and $\rm 10.0<log_{10}M_* < 10.5$
    (blue), $\rm 10.5<log_{10}M_* < 11.0$ (red) and $\rm 11.0<log_{10}M_* < 11.5$
    (mauve).} The horizontal dashed lines show the predicted star-formation rate for
    a galaxy on the galaxy `main sequence' at a redshift of 3 with $\rm log_{10} M_* = 10.25$ (blue),
    $\rm 10.5$ (red) and $\rm 11.25$ (mauve).
    }
	\label{fig:morphologies}
\end{figure}

\subsection{Is quenching already happening?}

Analysis of the morphological classifications of the COSMOS-Web galaxies
suggests that the familiar morphological types - the Hubble sequence -
came into existence at $\rm z \sim 4$ \citep{huertas2025}. In the universe today, there
is a clear progression along the Hubble sequence from the irregulars
and disk-dominated systems, which are generally gas-rich and with
high star-formation rates, to the bulge-dominated and spheroidal galaxies at the
other end of the sequence, which
contain less gas and have lower star-formation rates
\citep{kennicutt1998,saintonge2022}.
There must therefore must be
some quenching process that has differentially reduced the
star-formation rate along the Hubble sequence.
An obvious question to ask is: when did this differential quenching process
start?

There is already some
indication in Figure 10 that at high redshifts and high stellar masses
the mean star-formation rate for the spheroids and the bulge-dominated
galaxies is lower than for the other two morphological
classes. In this section we try to quantify this difference. We focus on the redshift range $\rm 2 < z < 4.5$, which
is roughly where analysis of the spectra of low-redshift massive ellipticals
implies most of the stars in these galaxies must have formed \citep{thomas2005,thomas2010}. We only consider galaxies
with $log_{10}M_* > 10.0$ because most galaxies with
lower stellar masses are irregulars.

Figure 12 shows the mean star-formation rate for the galaxies in this redshift
range for the four morphological classes, arranged in order along the Hubble sequence.
The figure shows that for each of the three mass intervals there is a clear progression along the Hubble sequence with the mean star-formation rate increasing
by a factor of $\simeq$4 from spheroids to irregulars. 

Figure 13 shows a similar
analysis for this redshift range
using the star-formation estimates from the UV-to-near-IR SEDs in
the COSMOS-Web catalogue. In this case, we have plotted histograms of the
specific star-formation rate (star-formation rate
divided by stellar mass - SSFR) for the four morphological classes, only including
galaxies with $log_{10}M_*>10.5$ because 
otherwise the histograms become dominated by the galaxies in the lowest of the
three mass bins shown in Figure 12.
This figure shows the same things as Figure 12. There is a steady increase in this
redshift range in the
median values of SSFR along the Hubble sequence. 

Both figures imply that even
in an epoch that was only shortly after the Hubble sequence came into existence,
differential quenching was already in progress.
This conclusion does not depend on the accuracy
of the morphological classifications because any classification errors will tend to
reduce any differences observed between the classes. 

We have also plotted in the spheroid panel
a histogram of the values of SSFR for the ellipticals in the Herschel Reference
Survey, a volume-limited survey of the nearby universe \citep{boselli2010,eales2017}.
The difference in SSFR between these low-redshift galaxies and most of the
spheroids in COSMOS-Web shows that even if quenching was already in action,
it had a long way to go.

\subsection{When did quenching start and how fast did it happen?}

Answers to both these questions are suggested by the form of the histograms
in Figure 13. The histograms of SSFR for both the irregulars and the disk-dominated
galaxies have clear peaks in this figure, which is similar to the expected
values for the `galaxy main sequence' \citep{popesso2023}. It seems a reasonable assumption that this would be where galaxies would be in the absence of
any quenching process. The histogram for the spheroids shows no peak at
all at this value of SSFR, suggesting that quenching must have started
either immediately after or even before the formation of the structures of the
spheroids.
The histogram for the bulge-dominated galaxies shows a slight peak and then
a long tail to lower values of SSFR, suggesting that quenching started
some time after the formation of the galaxy's structure.

The form of the spheroid histogram also suggest that the quenching process
is not very fast. Suppose, for example, the quenching occurs as the result of the
formation of a quasar at the centre of the galaxy some time after
the formation of its spheroidal structure, with the gas then driven out
by the pressure of the radiation from the quasar \citep{fabian2012,bollati2024}, rapidly shutting off star formation.
In this case, we would expect to see a histogram with two peaks rather than
the flat histogram we actually observe.

We have estimated a timescale for the quenching of the star formation in
the spheroidal class using the following very simple model.
We make the assumption that the moment the structure of the spheroid is
formed the quenching process starts. We assume that the spheroid initially has
a specific star-formation rate ($\rm SSFR_{init}$) that puts it on the
galaxy `main sequence' \citep{popesso2023}. We make the additional assumption,
which is suggested by the flat histogram, that the quenching
occurs by some kind of
`starvation' process in which the 
gas supply is shut off and the star-formation rate then falls as the result of the
gradual depletion of the galaxy's gas reservoir by the formation of
new stars. With this assumption,
the star-formation rate, $SFR(t)$, then depends on time in the following way:

$$
SFR(t) = SFR_{init} \times exp{(-t/\tau_{dep})}
$$

\noindent in which $SFR_{init}$ is the initial star-formation rate, $t$ is the time
since the gas supply was shut off, and $\tau_{dep}$ is the depletion time, the
one adjustable parameter in our very simple model. We make the
simplifying assumption that the stellar mass of the galaxy does not change
significantly after the quenching starts, so we can also write:

$$
SSFR(t) = SSFR_{init} \times exp{(-t/\tau_{dep})}
$$

\noindent Our final assumptions are that massive spheroids are
produced at a constant rate over the redshift range $2<z<4.4$ and that $SSFR_{init}$ remains constant, with a value of -8.9, over this redshift range. We can safely assume that as the star-formation rate in the galaxy falls,
it will still remain above the COSMOS-Web detection limit since the galaxies in
Figure 13 are well above the COSMOS-Web stellar mass limit \citep{shuntov2025a}.

The lines in Figure 13 shows the predictions of the model 
for three different depletion times. The model that provides
the best fit to the histogram has a depletion time
of $\rm \tau_{dep} = 10^{8.2}\ years$. Shorter depletion times produce too many galaxies with low
values of SSFR and vice versa. This is a shorter depletion time than that found
for low-redshift starbursts ($\rm \tau_{dep} \simeq 10^{8.5}\ years$ \citet{kennicutt2021}), but longer than the typical depletion time for
SMGs \citep{dye2015,dye2022}.

\begin{figure} 
	\includegraphics[width=\columnwidth]{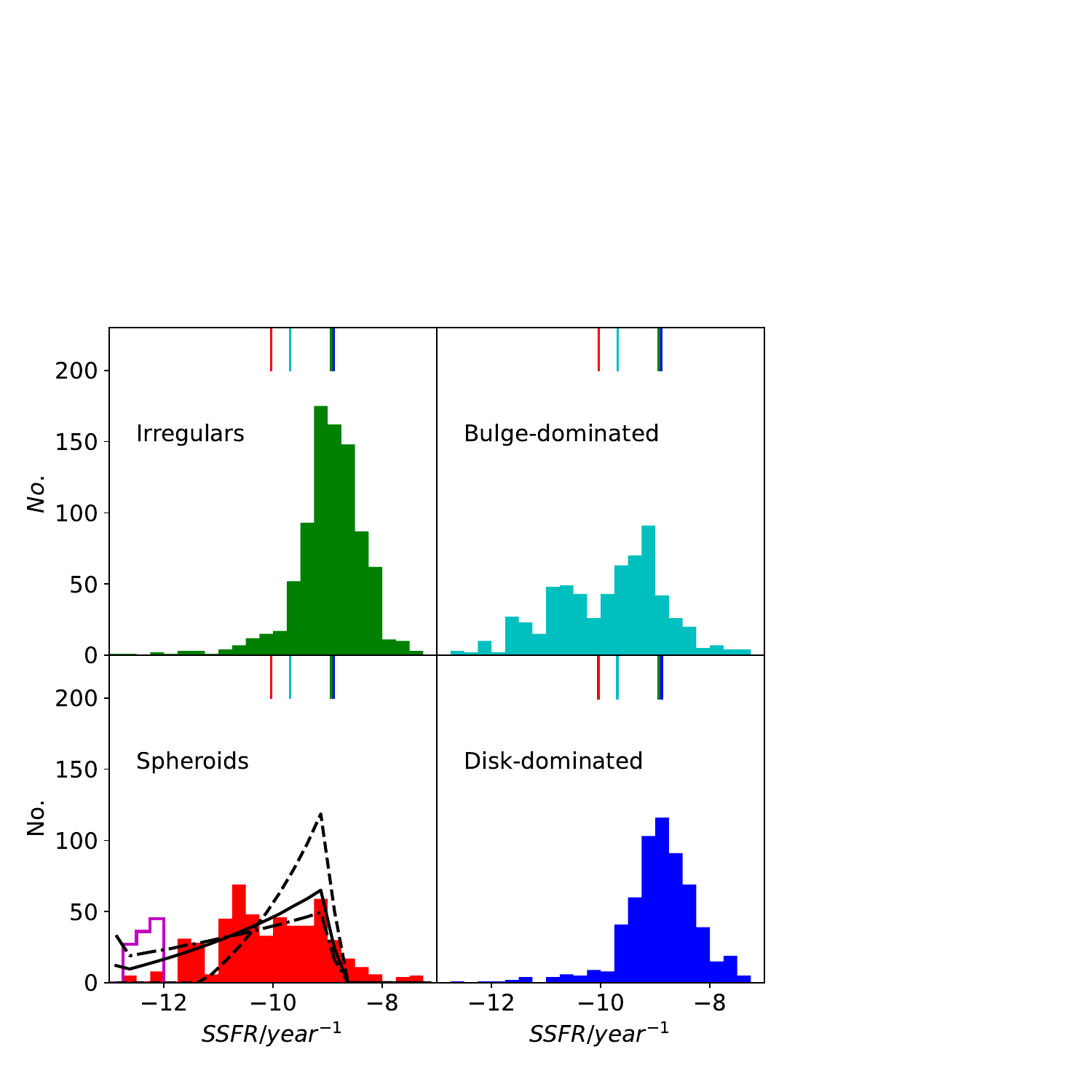}
	\caption{Histograms of the specific star-formation rate
    for the four morphological classes in the redshift range
    $\rm 2 < z <4.5$ and with $\rm 11.5 > log_{10}M_* > 10.5$.
    In this case, the star-formation rates are the values
    estimated from UV-to-near-IR photometry in the COSMOS-Web catalogue (\S 2.2).
The vertical lines show the median values of SSFR for the four classes. The
lines for the irregulars and the disk-dominated are so close as to
be indistinguishable. The lines in the spheroid panel shows
the prediction  of the quenching model described in \S 4.3
for gas depletion times of $\rm 10^{8}\ years$ (dot-dashed),
$\rm 10^{8.2}\ years$ (solid) and $\rm 10^{8.5}\ years$ (dashed).
The unfilled histogram in the spheroid panel
show the distribution of SSFR for the elliptical galaxies
in the Herschel Reference Survey \citep{devis2017}
}
	\label{fig:morphologies}
\end{figure}

    \begin{figure}
	\includegraphics[width=\columnwidth]{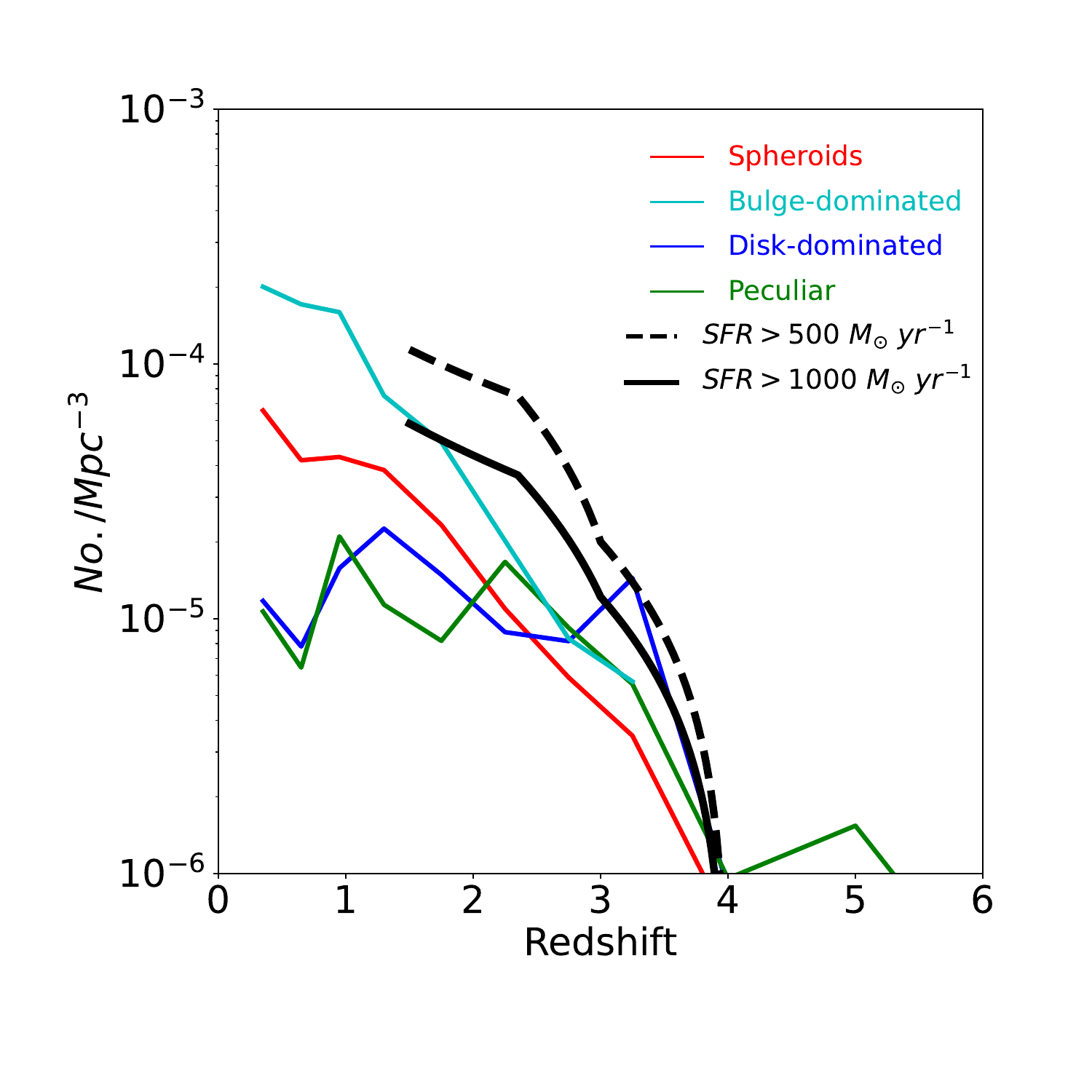}
	\caption{The space-density of galaxies in the mass range $\rm 11.0 < log_{10}M_*
    < 11.5$ with
    irregular morphologies (green), disk-dominated morphologies
    (blue), bulge-dominated morphologies (cyan) and
    spheroidal morphologies (red), calculated from the COSMOS-Web stellar mass function \citep{huertas2025}. The solid and dashed blacks lines shows the predictions
    of the model for the production of submillimetre galaxies (SMGs) in the same
    range of stellar mass for a lower star-formation limit of 500 and 1000 $\rm M_{\odot}\ yr^{-1}$, respectively.}
	\label{fig:morphologies}
\end{figure}

\subsection{Do massive spheroidal and bulge-dominated
galaxies pass through an SMG phase?}

In S4.1 we noted that the SMGs are not necessarily a distinct population from other disk-dominated and irregular galaxies, but might simply be extreme members
of this population with the highest star-formation rates. Nevertheless, there
is evidence that massive elliptical galaxies did pass through an early phase in which their star-formation rates were as high as those for the SMGs \citep{thomas2005,thomas2010}. Therefore, even if the SMGs are not a separate
population, it is still valid to ask whether
there are enough SMGs for them to represent this early phase in the lives
of massive early-type galaxies.

Previous studies have shown that there are enough SMGs for them to be the ancestral population of massive quiescent galaxies at $z>1$
\citep{valentino2020,long2023}. These studies, however, did not have access to the JWST morphological classifications. In this
section, we construct a model to test whether the rate of growth 
in the number-density of massive
spheroidal and bulge-dominated galaxies as found in COSMOS-Web is what one would expect from the
properties of the SMG population.

Figure 14 shows the space-density of galaxies
with stellar masses in the range $\rm 11.0 < log_{10}M_* < 11.5$ plotted against
redshift
for the four different morphological classes, which we have calculated from the
stellar mass functions for these classes \citep{huertas2025}. The figure
shows that the relationship is steeper for the spheroidal and bulge-dominated
galaxies than for the irregulars and disk-dominated galaxies. (This figure is
also prima-facie evidence that massive disk-dominated and irregular galaxies must be morphologically transformed
into something else because otherwise the relationships for these morphological types would be steeper than horizontal because of the continual formation of new stars out of gas within existing galaxies.)

Our model is based on the assumption that SMGs reach stellar
masses of $ log_{10}M_* > 11.0$ before the star formation is quenched. This
seems a reasonable assumption since a large fraction of SMGs already have
stellar masses above this value \citep{simopson2017,eales2024a}. We start with the
estimates of the space-density of SMGs as a function of bolometric luminosity,
$\phi(L_{bol})$, estimated by \citet{dud2020} for the redshift ranges
$z<2.35$, $2.35<z<3.0$ and $z>3.0$. For any galaxy
we can estimate the value of its star-formation rate from its bolometric luminosity using the calibration
given in \citep{kennicutt2012}, which allows us to convert
the space-density of SMGs as a function of bolometric luminosity
into the space-density of SMGs as a function of star-formation rate.

The time required for an SMG to reach 
a stellar mass of $\rm 10^{11}\ M_{\odot}$ is given by:

$$
\tau_{lifetime} = {10^{11} \over SFR}\ yrs
$$

\noindent in which $SFR$ is its star-formation rate. The number of galaxies with stellar masses above this mass ($N > 10^{11}\ M_{\odot}$) generated by the SMG population
is given by:

$$
N(t) = \int_{t_{start}}^t \int_{SFR_{min} (L_{bol})}^{\infty} {\phi(L_{bol})  \over \tau_{lifetime} } dL_{bol} dt
$$

\noindent in which $t_{start}$ is the cosmic time at which the SMG epoch started. 
We used this equation to predict the increase in the number of massive
galaxies over the redshift range $1.5<z<4$, the redshift range in which SMGs are
mostly found, using the luminosity functions mentioned above
for two different lower limits for the star-formation rate: 500 and 1000 $M_{\odot}\ yr^{-1}$

Figure 14 shows the results. The predictions for the model are very similar to
the COSMOS-Web results for the growth in the number-density of massive spheroidal
and bulge-dominated galaxies.

The figure shows that the growth in the number-density
of massive spheroidal and bulge-dominated galaxies is roughly what one would expect
if all these galaxies pass through an SMG phase. 
The space-density of massive spheroidal and bulge-dominated galaxies continues to climb at lower redshifts, which our model does not explain because of the lack of SMGs at lower redshifts. The production of massive early-types galaxies, spheroids and bulge-dominated galaxies, is therefore
not solely the result of the conversion of this seed population.

\subsection{The evolution of dust in the universe}

In this section we try to understand the evolution in the mean
dust density in the universe shown in
Figure 8.  We will also try to understand why the ratio of dust mass to
stellar mass increases with redshift (Figure 7).

There are now a very large number of sophisticated models for the evolution of dust in
galaxies (e.g. \citet{morgan2003,asano2013,zhukovska2016,popping2017,mckinnon2017,aoyama2018,vijayan2019}).
These all attempt to incorporate in different ways the complex physics of how dust is formed in stars and then distributed
into the ISM. \citet{magnelli2020}, who found a similar relationship
between mean dust density and redshift to ours (Figure 8), tested some of these
models, finding that they were not able to reproduce the observed
relationship \citep{popping2017,aoyama2018,li2019,vijayan2019}.

In this section, we try to make a zeroth-order attempt
to understand the relationships in Figures 7 and 8 by using a simple model that
cuts out all the complex, and uncertain, physics. 
In our earlier paper \citep{eales2024b}, we pointed out the
similarity of the star-formation history of the universe and the
relationship between the mean dust density and redshift, which
was also noticed by \citet{magnelli2020}, with the only difference 
being that at $\rm z > 2$ the mean dust density
falls more rapidly with redshift than the mean star-formation rate.
One of the advantages of our modelling approach is
that, by stripping away the complexity, it makes it possible to get an
understanding of the relationship between the star-formation history
and the evolution of the dust.

The first assumption of the model is that a constant fraction
of the metals in the ISM in a galaxy are incorporated in dust grains. There is evidence,
from observations of galaxies in the Local Group and damped Lyman alpha systems,
that this is true for
galaxies with high metal abundances, with about half the metals being incorporated 
in dust \citep{decia2016,romanduval22,romanduval22b}.
The slope of the relationship between the dust-to-gas ratio and metal abundance is less than
one \citep{romanduval22}, so the fraction of
the metals incorporated in the dust grains will be lower 
in galaxies with low metal abundance, but in this initial simple model we assume it is a 
constant. The advantage of this assumption
is that we can avoid the poorly understood physics of how the metals are incorporated in the
dust \citep{zhukovska2016} and use standard chemical evolution models to model the evolution of the dust.

The chemical evolution model we use is a simple analytic model, which relates the evolution
of the mean metal abundance in a galaxy to the star-formation rate, the inflow of gas and
the outflow of gas. We refer the reader to \citet{edmunds90a} for a full description
of the model, but we give here its main points.

The model is based on three additional assumptions. The first is that the
mixing of metals within a galaxy is instantaneous, 
so the chemical evolution of the galaxy can be fully represented
by the evolution in the mean metal abundance.
The second is that
the yield - the mass of metals produced by 1 solar mass of newly formed stars - is
a constant. The third is the assumption of `instantaneous recycling', in
which the metals that will eventually be produced by a generation of stars are, in the
model, deposited in the ISM the moment the stars are born. This assumption becomes
dubious at very high redshifts when the ages of stars are similar to the age of the universe
\citep{eales2024a}, but it is a powerful assumption because it makes it possible to
relate the metallicity at any time (and the amount of dust) to the star-formation
rate at the same time.

In the following equations, $g$ is the mass of gas in a galaxy, $s$ is the mass of stars in
the galaxy, $R$ is the fraction of the mass of a generation of stars that
is ultimately returned to the ISM rather than being locked up in stellar remnants and very
long-lived stars, $y$ is the yield, the mass of metals produced by one solar mass of newly
formed stars, and $Z$ is the metal abundance of the gas. 
We assume that the outflow of gas from the galaxy is given
by $\Lambda$ times the star-formation rate and the inflow of gas (assumed to contain
no metals) is allowed to have an arbitrary dependence on time given  by $\Phi(t)$. 

The two basic equations from the chemical evolution model \citep{edmunds90a} are
as follows. The change in the mass of metals in the gas is given by:

\begin{equation}
    d(Zg) = R y ds - (1-R) Z ds - \Lambda Z ds
	\label{eq:chem-ev 1}
\end{equation}

\noindent In this equation, the first term on the right-hand side is the mass of metals injected into the ISM
by newly formed stars. The second term is the mass of metals locked up in stellar remnants and long-lived
stars. The third term is the mass of metals ejected from the galaxy. The change in the mass of gas in
the galaxy is given by:

\begin{equation}
    dg = \Phi dt - (1-R) ds - \Lambda ds
	\label{eq:chem-ev 2}
\end{equation}

We can now manipulate these equations into a form that is useful for a model for the evolution of the
dust. Reorganizing the second equation a little gives:

\begin{equation}
    \Phi(t) = {dg \over dt} + \Lambda {ds \over dt} + (1-R) {ds \over dt}
	\label{eq:chem-ev 3}
\end{equation}

\noindent Combining the two equations yields:

\begin{equation}
    {dZ \over dt} = {Ry \over g} {ds \over dt} - {Z \Phi \over g}
	\label{eq:chem-ev 4}
\end{equation}

\noindent We can also write down a third equation if we assume that the star-formation rate in a galaxy is
proportional to the mass of gas it contains \citep{kennicutt1998,kennicutt2012}:

\begin{equation}
    {ds \over dt} = {g \over \tau_{dep}}
	\label{eq:chem-ev 5}
\end{equation}

\noindent in which $\tau_{dep}$ is the depletion time of the gas in the galaxy.

This simple model of the evolution of the metal abundance in a galaxy can be simply
adapted to model the mass of dust in a galaxy with the assumption that the mass of dust
is given by:

\begin{equation}
    M_d = \epsilon_d Zg
	\label{eq:chem-ev 6}
\end{equation}

\noindent in which $\epsilon_d$ is the fraction of the metals locked up in dust grains.

This simple model has been used before to model
the evolution of the dust in individual galaxies \citep{eales96,eales97,edmunds98}. In this paper, we extend 
it to the
universe as a whole. The necessary assumptions are virtually the same. 
In the case of an individual
galaxy, although though there are radial metallicity gradients in galaxies \citep{maiolino2019},
it is necessary to assume that the 
chemical evolution of a galaxy can be fully described by the
evolution of the mean metal abundance. In the case of the universe, one has to make effectively the same assumption - that although the metal
abundances of the galaxies in a volume of space are certainly very different, the chemical evolution of the universe can be
usefully modelled by the change in its mean metal abundance. 
In this extension of the model
to the universe as a whole, $g$ is now the mean density of gas in the universe, $s$ is the mean stellar
density, $\Phi$ is the mean flow of pristine gas into galaxies, $\Lambda {ds\over dt}$ is the mean outflow of
gas from galaxies into the intergalactic medium, and $M_d$ is the mean density of dust.

We implemented the model in the following way. We used the analytic expression for the
star-formation history of the universe \citep{madau2014} and equation 12 to deduce the relationship between
the mean gas density and redshift, using a depletion time of $\rm 10^{8.5}\ years$,
a value appropriate for galaxies with high star-formation rates \citep{kennicutt2021}.
Many authors have argued that the depletion time varies with redshift
\citep{scoville2016,scoville2017,tacconi2020,scoville2023}, so 
our assumption of a constant value could be wrong, but it is useful
simplification, and we also argue in the next section that the evidence for a
changing depletion time may be at least partly caused by changes in the gas-to-dust
ratio.

With the star-formation history and the relationship between gas density and redshift,
we are then able to use equation 10 to calculate the time dependence of the rate of
inflow of pristine gas, $\Phi(t)$. We then integrate equation 11 to derive the evolution
of the mean metal abundance of the universe, starting with $Z=0$ at $t=0$, and then use equation 13
to derive the evolution of the mean dust density. In our model we use a value for $\epsilon_d$
of 0.52 \citep{romanduval22},  and values for $R$ and $y$ of 0.38 and 0.04, respectively, which are appropriate
for a Chabrier initial mass function \citep{eales2024a}.

Figure 15 shows the results for $\Lambda=0$, $\Lambda=1$ and $\Lambda=2$. The model agree well with the
mean dust density calculated with {\bf COLD} and
{\bf STAR-FORMER} but predicts more
dust in the early universe than the estimates
for {\bf EVOLVE}. Of the three dust-evolution
models, the one that agrees best with the data (as calculated with the first two temperature
models) is the model with outflow parameter $\Lambda=1$.

\begin{figure}
	\includegraphics[width=\columnwidth]{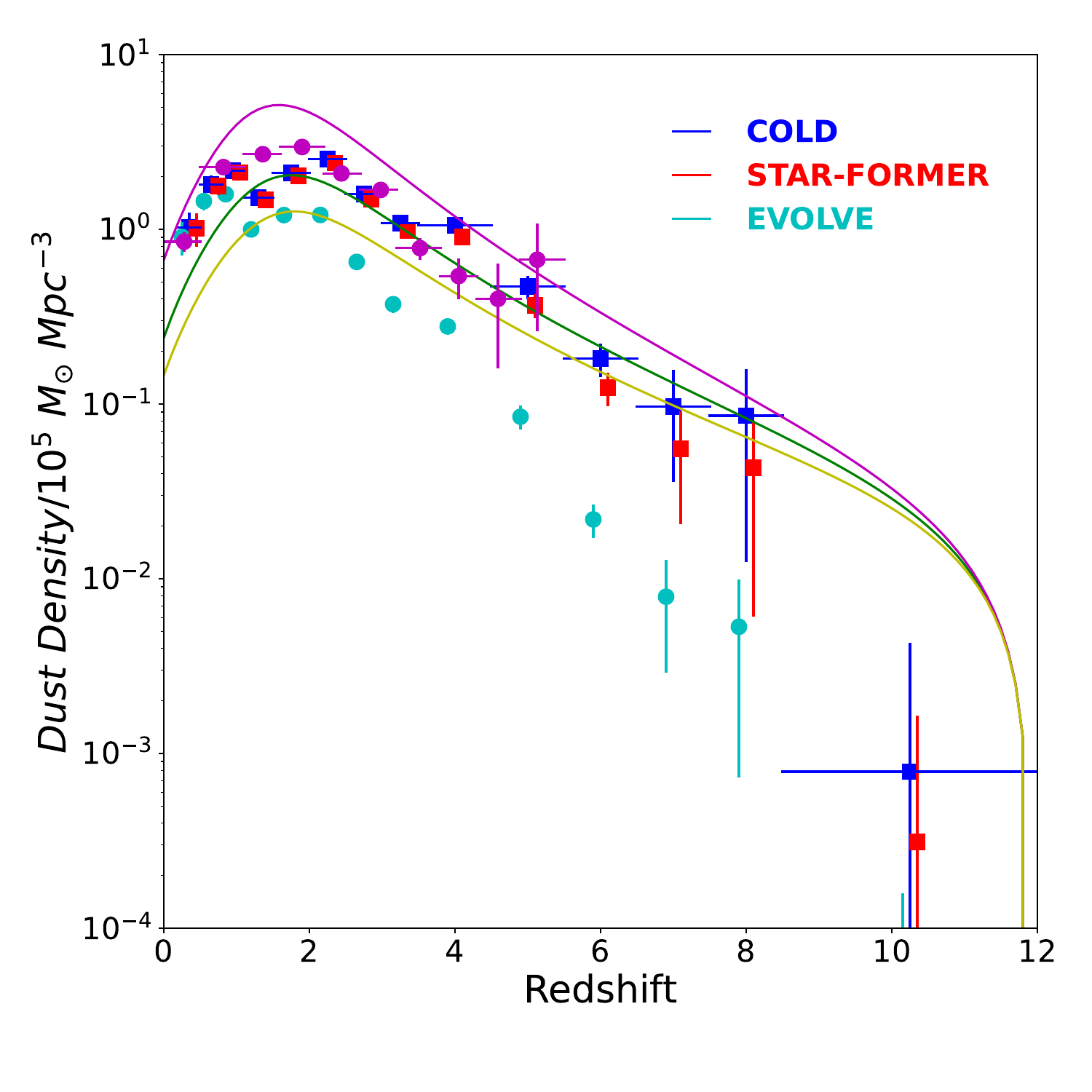}
	\caption{The same as Figure 8 with the addition of the predictions
    of our model for the evolution of cosmic dust (see text). The purple line
    is for a closed-box model, the green line for an outflow model
    with $\Lambda=1$ and the yellow line for an outflow model with
    $\Lambda=2$.}
    \label{fig:dust density}
\end{figure}

We can extend this model to make a prediction for how the
ratio of dust mass to stellar mass should change with redshift.
We calculated the relationship between the mean stellar density
and redshift by integrating the analytic expression for the star-formation history \citep{madau2014}. It is then a simple matter to calculate
how the ratio of the dust mass to stellar mass should depend
on redshift from the relationship between mean dust density
and redshift calculated above. The predictions
for the three dust-evolution models are shown in Figure 7. In this case, the agreement between the
model predictions and the observations is not so compelling, but the observed increase in the mass ratio with
redshift is clearly seen in all three models. 

This simple model allows an intuitive understanding of why
the ratio of the mass of dust to stars is greatest
at early times. Equation 8 shows that the change in the mean metal density
(and in our model the mean dust density) is comprised of three components.
One of these, the formation of metals in stars, leads to an increase in the mean metal density.
The two others - outflows and the consumption of metals when new stars are formed (astration) -
lead to a reduction in the mean metal density.
The two latter terms are proportional to the metal abundance in the gas.

Since we assume that the metal production is proportional to the star-formation
rate, the mass of metals (and dust) in the ISM and the mass of
stars will initially increase at the same rate when there are no metals
in the ISM. But as the metals build up in the ISM and depletion from outflows
and astration become significant, the mass of metals and dust in the ISM will
increase at a slower rate than the mass of stars. Therefore, it is
not surprising that the model predicts that the ratio of the mass of dust
to the mass of stars should be greatest at early times when not much
dust production has occurred. We note the
caveat that this is also when the limitations of our simple
model will be most apparent, since we have made the useful simplifying assumption
that the dust is instantly made when the stars are formed and this will
be a poor approximation when the age of the universe is similar to the
lifetimes of stars.

\subsection{The evolution of dust-traced gas in the universe}

Dust has some major advantages over CO for estimating the amount of molecular
gas in a galaxy \citep{magdis2012,eales2012,scoville2014,tacconi2020,dunne2022}. In this section, we use the estimates of the cosmic dust density (\S3.5) 
to estimate how the mean density of dust-traced gas changes with
redshift. 

We assume, fairly arbitrarily, that the
average gas-to-dust ratio in galaxies today is 100. Many recent JWST studies have shown
that the typical metal abundance in galaxies decreaes with increasing redshift (\citet{ellis2025} and references
therein). There is now a reasonable consistent picture that
the metal abundance in galaxies falls by 0.8 dex from $\rm z=0$ to $\rm z=3$, with a smaller
decline of 0.2 dex from $\rm z=3$ to $\rm z=10$ \citep{ellis2025}. We assume the following relationships
between metal abundance, $A$, and redshift that are consistent with these conclusions:

\begin{equation}
    log_{10}A = log_{10} A(z=0) - 0.267 z\ \ \ \ \ \ 0<z<3
	\label{eq:dust density 2}
\end{equation}

\begin{equation}
    log_{10}A = log_{10} A(z=0) -0.714 - 0.0286 z \ \ \ \ \ \ 3 < z < 10
	\label{eq:dust density 2}
\end{equation}

\noindent Although the metal abundance also depends on the mass of the galaxy \citep{maiolino2019}, we haven't got any straightforward way of including the mass dependence in our anyalysis, and anyway the mass dependence is poorly known at the highest redshifts. Therefore, in this
initial attempt to make a correction for the evolution in the
gas-to-dust ratio, we have not included any dependence on stellar mass.

On the assumption that the dust-to-gas ratio is proportional to the metal abundance \citep{decia2016,romanduval22},
we used these relationships to calculate how the dust-to-gas ratio changes with redshift. 
With these relationships, the gas-to-dust ratio increases from 100
at $z=0$ to $\simeq$720 at $z=5$ and $\simeq$1000 at $z=10$.

We use this relationship between gas-to-dust ratio and redshift
to calculate the gas density from the estimates of the mean dust density
for the three temperature models (\S 2.7).
Figure 16 shows the result. The dashed line and the coloured band
shows a recent estimate, and its uncertainty, for the evolution of the
CO-traced gas \citep{peroux2020}. The shape of the two relationships,
one for CO-traced gas and one for dust-traced gas, are very similar.
There is a vertical offset, but this is not surprising given our
fairly arbitrary choice of gas-to-dust ratio and that our estimate
also depends on our choice of dust-mass-opacity coefficient. There is
a similar uncertainty in the scaling of the relationship for CO because
of the uncertainty in the CO calibration factor \citep{bolatto2013,dunne2022}.\footnote{In our previous paper \citep{eales2024b}, we used a slightly different method so
that the calibrations of the dust and the CO were self-consistent. We
still ended up with a vertical offset.}

We have overplotted 
a relationship derived for the star-formation history of the universe
\citep{madau2014}. We have scaled the two vertical axes so that
our relationship between dust-traced gas and the star-formation
history coincides as far as is possible. The scaling corresponds
to a gas depletion time of:

$$
\tau_{dep} = 6.3\times 10^8 \left( {gtd \over 100} \right)\ years
$$

\noindent in which $gtd$ is the gas-to-dust ratio. With our fairly arbitrary assumption that the gas-to-dust ratio is 100, our estimate
of the gas depletion time is similar to the values found by \citet{tacconi2020} at $z\simeq1$.

Although the scaling has been chosen in this way,
the forms of the star-formation history and the relationship
between dust-traced gas and redshift 
are remarkably similar. This is only true, though, for {\bf COLD} and
{\bf STAR-FORMER}. The similarity of the relationships, at least
with our preferred {\bf STAR-FORMER} temperature model, is quite
remarkable. It suggests that the gas depletion time, or conversely
the star-formation efficiency, is independent of redshift.
This conclusion does differ from claims in the literature that
the depletion time decreases with redshift \citep{scoville2016,scoville2017,tacconi2020,scoville2023}.
The difference arises because we have assumed that the gas-to-dust
ratio increases with redshift whereas Scoville and collaborators,
whose work is most easily comparable to ours, assume that the gas-to-dust ratio
is a constant.

\begin{figure}
	\includegraphics[width=\columnwidth]{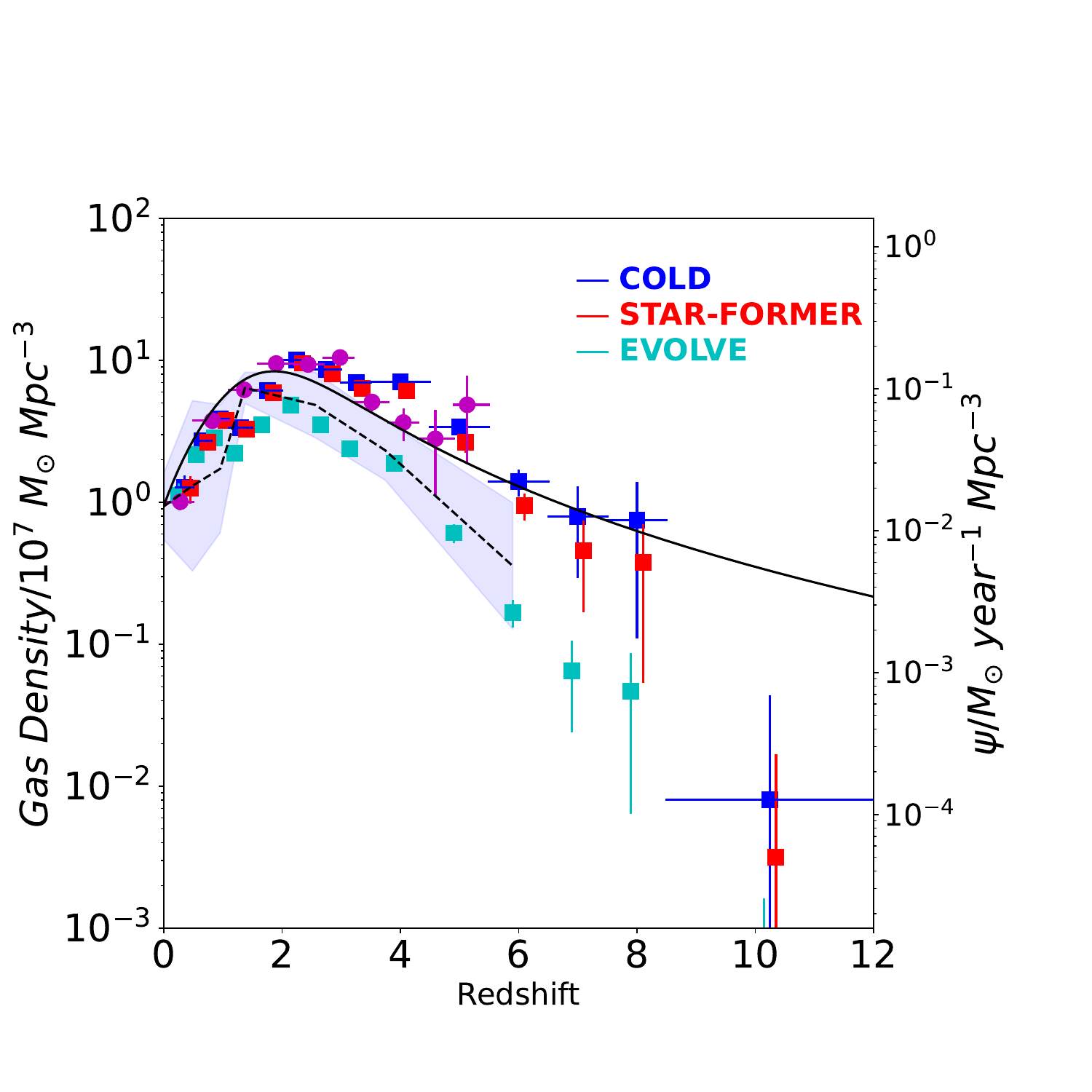}
	\caption{The mean density of dust-traced gas versus redshift for {\bf COLD} (blue),
    {\bf STAR-FORMER} (red) and {\bf EVOLVE} (cyan) and from our previous study (purple).
     The solid line
 shows the `star-formation history of the universe', an analytic
 relationship derived
 from empirical estimates of the global star-formation rate \citep{madau2014}.
 The star-formation-density axis has been scaled so that the
 star-formation history coincides with our measurements of the
 gas density, with the scaling corresponding to
 a gas depletion time of $\rm 6.3\times10^8\ years$.
 The dashed line and the coloured band show the best current estimates
 and their uncertainty for the CO-traced gas \citep{peroux2020}.}
	\label{fig:dust density}
\end{figure}

\section{Conclusions} 

We have used two different stacking techniques to measure the mean submillimetre (850 $\mu$m) 
flux density
for galaxies in the COSMOS-Web catalogue separated into samples by stellar mass, redshift and
morphological class. The two techniques gives similar results, showing a gradual increase of
submillimetre flux density with redshift, and at each redshift a positive correlation between flux
density and stellar mass.

We converted the submillimetre flux densities into dust masses and bolometric
dust luminosities using three different
temperature models. We have tried to avoid the temperature bias
that arises from fitting a single modified blackbody to a galaxy SED by using  
a more sophisticated model based on the SEDs of real star-forming
galaxies. We have
included the two other models to show the effect on our results of the range
of dust temperatures in the literature. We have used our stacking results and
temperature models to obtain the following results:

\begin{enumerate}

\item 
We have used our stacking results and temperature models to determine
the mean bolometric dust luminosities for each sample and thus estimate
the mean star-formation rate for each sample. 
In contrast to the situation for submillimetre galaxies, we find no evidence
for any systematic difference between our estimates of the star-formation rate
and the mean of the estimates in the COSMOS-Web catalogue, which were estimated
from fitting stellar population synthesis models to the
UV-to-near-IR SEDs.

\item
The mean star-formation rates for all morphological classes increase with redshift
roughly following the relationship found for the `galaxy main sequence'
\citep{popesso2023},
reaching star-formation rates 
for the most massive galaxies ($M_* \sim 10^{11}\ M_{\odot}$)
of $\rm \gtrapprox 80\ M_{\odot}\ yr^{-1}$ at $z>2$.

\item In the redshift range $2 < z < 4.5$ the mean
star-formation rate decreases
along the Hubble sequence from $\rm \simeq 280\ M_{\odot}\ yr^{-1}$ for irregular galaxies at one end to $\rm \simeq80\ M_{\odot}\ yr^{-1}$for spheroids at 
the other end. Since this only shortly after the emergence of the Hubble sequence
($z \sim 4$ - \citet{huertas2025}), we conclude that quenching 
was already underway, or even before, the structures of the galaxies were
formed.

\item 
We show that star-formation rates for the high-redshift spheroids
can be reproduced by a `starvation' quenching model with
a depletion time of $\rm \simeq10^{8.2}\ years$. 

\item 
We use the COSMOS-Web stellar mass functions \citep{huertas2025} to investigate the growth of massive
($>10^{11}\ M_{\odot})$ galaxies of different morphological types over the
redshift range $0<z<4$. The number-density of massive irregular and disk-dominated
galaxies is fairly constant while the number-density of bulge-dominated and
spheroidal galaxies increases steadily over this period. We
construct a model based on the assumption that the seed population of massive spheroidal and bulge-dominated
galaxies are `submillimetre galaxies' and show that this model can reproduce the growth in the number-density of these galaxies over the redshift range $1.5 <z < 4$.
It does not explain the continued growth at lower redshifts where there are few SMGs, which shows that the production of massive early-type
galaxies is not entirely the result of the conversion of this seed population.

\item 
We have used our stacking results
to show that the ratio of dust mass to stellar mass in galaxies increases with redshift out
to $z \sim 8$.

\item 
We have combined our stacking results with the COSMOS-Web stellar mass functions to determine the relationship between the mean density of dust and redshift in the range $\rm 0 < z <12$, extending the work in our earlier paper \citep{eales2024b}.

\item We have constructed a model for the global evolution of dust in the universe
based on the empirical star-formation history of the universe and on previous models for the evolution of the metals in galaxies. We show that with a mean
gas outflow rate equal to the star-formation rate this model reproduces our
observed relationship between mean dust density and redshift remarkably well. The model also explains why the mean ratio of dust mass to stellar mass
increases monotonically with redshift.

\item 
We use our dust-redshift relationship and assumptions about
the evolution of the mean metal abundance in the universe based on recent JWST observations to derive the relationship between the mean density of molecular
gas and redshift. The shape of this relationship at the lower redshifts
($\rm z<6$) is similar to the relationship derived using CO rather
than dust as the tracer, and the offset can easily be explained
in the uncertainties in calibrations of the CO and dust methods.

\item
The relationship we derive between gas density and redshift is, for a gas depletion time
of $\rm 6.3\times10^8\ years$, remarkably similar to the star-formation
history of the universe.
The similarity of the two relationships suggests that the
global gas depletion time has remained roughly constant
over the history of the universe.

\end{enumerate}

\section*{Acknowledgements}

We thank Hiddo Algera, Richard Ellis, Karl Glazebrook and Ryley Hill for comments
on the manuscript. We are particularly grateful to Ian
Smail for helping us get rapid access to the S2COSMOS dataset and for
his comments on the manuscript, which significantly improved the paper.
We are also grateful to an anonymous referee for their careful reading
of a very long paper and for many comments which significantly improved it.
This work was started while SAE was on a visit to Australia in summer 2025
funded by the Taith fund, Wales' international exchange programme. He thanks Luca Cortese
and Karl Glazebrook for helping to arrange his visits, respectively, to the International Centre for Research in Radio Astronomy and Swinburne University. SAE and MWSL are grateful to the Science and Technology Facilities Council (consolidated grant ST/K000926/1) for financial support.

\section*{Data Availability}

The programs used in this investigation are available on Github (https://github.com/steveales/COSMOS-Web), with the
exception of the SIMSTACK program, which can be found at https://github.com/marcoviero/simstack3/tree/main/viero22.



\bibliographystyle{mnras}
\bibliography{example} 




\appendix

\section{Flux Densities and Morphological Classification}

The flux densities measured with SIMSTACK shown in Figure 2 are listed
in Tables A1 and A2. The morphological breakdown of our redshift and stellar
mass samples is given in Tables A3 and A4.

\begin{table*}
	\centering
	\caption{Mean Flux Densities (SIMSTACK) }
	\label{tab:example_table}
	\begin{tabular}{llllll} 
		\hline
		$\rm z_{low}$ & $\rm z_{up}$ & $\rm log_{10}(M_{low}/M_{\odot})$ & $\rm log_{10}(M_{up}/M_{\odot})$ & $\rm <S_{850\ \mu m}>/mJy$ &
        $\rm \sigma/mJy$\\
		\hline
    0.2 & 0.5 & 9.0 & 9.5 & 0.030 &    0.024 \\ 
  0.2 & 0.5 & 9.5 & 10.0 & 0.014 &   0.029 \\ 
  0.2 & 0.5 & 10.0 & 10.5 & 0.048 &   0.040 \\
  0.2 & 0.5 &  10.5 & 11.0 & 0.41 &    0.05 \\
  0.2 & 0.5 & 11.0 & 11.5 & 0.24  &  0.07 \\ 
0.5 & 0.8 & 9.0 & 9.5 & 0.037 & 0.015 \\
0.5 & 0.8 & 9.5 & 10.0 & 0.11 & 0.02 \\
0.5 & 0.8 & 10.0 & 10.5 & 0.13 & 0.02 \\
0.5 & 0.8 & 10.5 & 11.0 & 0.17 & 0.03 \\
0.5 & 0.8 & 11.0 & 11.5 & 0.08 & 0.06 \\
0.8 & 1.1 & 9.0 & 9.5 & -0.00 & 0.01 \\
0.8 & 1.1 & 9.5 & 10.0 & 0.09 & 0.01 \\
0.8 & 1.1 & 10.0 & 10.5 & 0.14 & 0.02 \\
0.8 & 1.1 & 10.5 & 11.0 & 0.13 & 0.02 \\
0.8 & 1.1 & 11.0 & 11.5 & 0.10 & 0.05 \\
1.1 & 1.5 & 9.0 & 9.5 & -0.02 & 0.01 \\
1.1 & 1.5 & 9.5 & 10.0 & 0.04 & 0.02 \\
1.1 & 1.5 & 10.0 & 10.5 & 0.16 & 0.03 \\
1.1 & 1.5 & 10.5 & 11.0 & 0.32 & 0.03 \\
1.1 & 1.5 & 11.0 & 11.5 & 0.52 & 0.07 \\
1.5 & 2.0 & 9.0 & 9.5 & 0.036 & 0.013 \\
1.5 & 2.0 & 9.5 & 10.0 & 0.070 & 0.017 \\
1.5 & 2.0 & 10.0 & 10.5 & 0.26 & 0.02 \\
1.5 & 2.0 & 10.5 & 11.0 & 0.38 & 0.03 \\
1.5 & 2.0 & 11.0 & 11.5 & 0.66 & 0.07 \\
2.0 & 2.5 & 9.0 & 9.5 & 0.027 & 0.016 \\
2.0 & 2.5 & 9.5 & 10.0 & 0.12 & 0.02 \\
2.0 & 2.5 & 10.0 & 10.5 & 0.39 & 0.03 \\
2.0 & 2.5 & 10.5 & 11.0 & 0.82 & 0.06 \\
2.0 & 2.5 & 11.0 & 11.5 & 1.45 & 0.15 \\
2.5 & 3.0 & 9.0 & 9.5 & 0.01 & 0.02 \\
2.5 & 3.0 & 9.5 & 10.0 & 0.17 & 0.03 \\
2.5 & 3.0 & 10.0 & 10.5 & 0.49 & 0.06 \\
2.5 & 3.0 & 10.5 & 11.0 & 0.73 & 0.08 \\
2.5 & 3.0 & 11.0 & 11.5 & 1.30 & 0.29 \\
3.0 & 3.5 & 9.0 & 9.5 & -0.03 & 0.02 \\
3.0 & 3.5 & 9.5 & 10.0 & 0.05 & 0.04 \\
3.0 & 3.5 & 10.0 & 10.5 & 0.34 & 0.05 \\
3.0 & 3.5 & 10.5 & 11.0 & 0.92 & 0.09 \\
3.0 & 3.5 & 11.0 & 11.5 & 1.40 & 0.19 \\
3.5 & 4.5 & 9.0 & 9.5 & 0.02 & 0.02 \\
3.5 & 4.5 & 9.5 & 10.0 & 0.18 & 0.03 \\
3.5 & 4.5 & 10.0 & 10.5 & 0.74 & 0.06 \\
3.5 & 4.5 & 10.5 & 11.0 & 1.01 & 0.11 \\
3.5 & 4.5 & 11.0 & 11.5 & 2.11 & 0.27 \\
4.5 & 5.5 & 9.0 & 9.5 & 0.05 & 0.03 \\
4.5 & 5.5 & 9.5 & 10.0 & 0.15 & 0.05 \\
4.5 & 5.5 & 10.0 & 10.5 & 0.65 & 0.15 \\
4.5 & 5.5 & 10.5 & 11.0 & 1.82 & 0.29 \\
4.5 & 5.5 & 11.0 & 11.5 & 3.1 & 0.7 \\
5.5 & 6.5 & 9.0 & 9.5 & -0.04 & 0.05 \\
5.5 & 6.5 & 9.5 & 10.0 & 0.18 & 0.10 \\
5.5 & 6.5 & 10.0 & 10.5 & 0.68 & 0.30 \\
5.5 & 6.5 & 10.5 & 11.0 & 2.42 & 0.52 \\
5.5 & 6.5 & 11.0 & 11.5 & 2.40 & 0.77 \\
6.5 & 7.5 & 9.0 & 9.5 & 0.03 & 0.06 \\
6.5 & 7.5 & 9.5 & 10.0 & 0.10 & 0.12 \\
6.5 & 7.5 & 10.0 & 10.5 & 0.31 & 0.26 \\
6.5 & 7.5 & 10.5 & 11.0 & 0.82 & 0.59 \\
6.5 & 7.5 & 11.0 & 11.5 & 1.57 & 0.80 \\
\hline
	\end{tabular}
\end{table*}

\begin{table*}
	\centering
	\caption{Mean Flux Densities (SIMSTACK) - continued }
	\label{tab:example_table}
	\begin{tabular}{llllll} 
		\hline
		$\rm z_{low}$ & $\rm z_{up}$ & $\rm log_{10}(M_{low}/M_{\odot})$ & $\rm log_{10}(M_{up}/M_{\odot})$ & $\rm <S_{850\ \mu m}>/mJy$ &
        $\rm \sigma/mJy$\\
		\hline
7.5 & 8.5 & 9.0 & 9.5 & 0.05 & 0.09 \\
7.5 & 8.5 & 9.5 & 10.0 & 0.06 & 0.11 \\
7.5 & 8.5 & 10.0 & 10.5 & 0.42 & 0.25 \\
7.5 & 8.5 & 10.5 & 11.0 & 1.15 & 0.59 \\
7.5 & 8.5 & 11.0 & 11.5 & 2.33 & 0.85 \\
8.5 & 12.0 & 9.0 & 9.5 & -0.13 & 0.11 \\
8.5 & 12.0 & 9.5 & 10.0 & -0.23 & 0.14 \\
8.5 & 12.0 & 10.0 & 10.5 & 0.04 & 0.22 \\
8.5 & 12.0 & 10.5 & 11.0 & 0.05 & 0.52 \\
8.5 & 12.0 & 11.0 & 11.5 & 0.64 & 0.85 \\
\hline
	\end{tabular}
\end{table*}

\begin{table*}
	\centering
	\caption{Morphological Composition }
	\label{tab:example_table}
	\begin{tabular}{llllllllll} 
		\hline
		$\rm z_{low}$ & $\rm z_{up}$ & $\rm log_{10}(M_{low}/M_{\odot})$ & $\rm log_{10}(M_{up}/M_{\odot})$ & Total$^a$ & spheroids$^a$ & disks$^a$ & irregualr$^a$ & bulge-dominated$^a$ & uc$^a$\\
		\hline
0.2 & 0.5 & 9.0 & 9.5 & 1798 & 65 (3\%) & 725 (40\%) & 809 (44\%) & 167 (9\%) & 32 (1\%) \\
0.2 & 0.5 & 9.5 & 10.0 & 1026 & 52 (5\%) & 522 (50\%) & 262 (25\%) & 178 (17\%) & 12 (1\%) \\
0.2 & 0.5 & 10.0 & 10.5 & 803 & 81 (10\%) & 278 (34\%) & 99 (12\%) & 336 (41\%) & 9 (1\%) \\
0.2 & 0.5 & 10.5 & 11.0 & 550 & 82 (14\%) & 98 (17\%) & 17 (3\%) & 350 (63\%) & 3 (0\%)\\
0.2 & 0.5 & 11.0 & 11.5 & 113 & 28 (24\%) & 8 (7\%) & 4 (3\%) & 73 (64\%) & 0 (0\%) \\
0.5 & 0.8 & 9.0 & 9.5 & 3837 & 277 (7\%) & 1142 (29\%) & 2139 (55\%) & 256 (6\%) & 23 (0\%) \\
0.5 & 0.8 & 9.5 & 10.0 & 2290 & 147 (6\%) & 875 (38\%) & 986 (43\%) & 261 (11\%) & 21 (0\%) \\
0.5 & 0.8 & 10.0 & 10.5 & 1574 & 151 (9\%) & 526 (33\%) & 408 (25\%) & 481 (30\%) & 8 (0\%) \\
0.5 & 0.8 & 10.5 & 11.0 & 933 & 133 (14\%) & 189 (20\%) & 87 (9\%) & 518 (55\%) & 6 (0\%) \\
0.5 & 0.8 & 11.0 & 11.5 & 210 & 45 (21\%) & 8 (3\%) & 8 (3\%) & 146 (69\%) & 3 (1\%) \\
0.8 & 1.1 & 9.0 & 9.5 & 6812 & 505 (7\%) & 1663 (24\%) & 4272 (62\%) & 335 (4\%) & 37 (0\%) \\
0.8 & 1.1 & 9.5 & 10.0 & 4263 & 350 (8\%) & 1241 (29\%) & 2302 (53\%) & 350 (8\%) & 20 (0\%) \\
0.8 & 1.1 & 10.0 & 10.5 & 2751 & 324 (11\%) & 694 (25\%) & 1139 (41\%) & 576 (20\%) & 18 (0\%) \\
0.8 & 1.1 & 10.5 & 11.0 & 1550 & 217 (14\%) & 281 (18\%) & 306 (19\%) & 737 (47\%) & 9 (0\%) \\
0.8 & 1.1 & 11.0 & 11.5 & 306 & 69 (22\%) & 23 (7\%) & 21 (6\%) & 190 (62\%) & 3 (0\%) \\
1.1 & 1.5 & 9.0 & 9.5 & 6190 & 318 (5\%) & 1827 (29\%) & 3666 (59\%) & 352 (5\%) & 27 (0\%) \\
1.1 & 1.5 & 9.5 & 10.0 & 3652 & 279 (7\%) & 1441 (39\%) & 1584 (43\%) & 331 (9\%) & 17 (0\%) \\
1.1 & 1.5 & 10.0 & 10.5 & 2206 & 506 (22\%) & 645 (29\%) & 647 (29\%) & 398 (18\%) & 10 (0\%) \\
1.1 & 1.5 & 10.5 & 11.0 & 1547 & 349 (22\%) & 394 (25\%) & 180 (11\%) & 621 (40\%) & 3 (0\%) \\
1.1 & 1.5 & 11.0 & 11.5 & 325 & 86 (26\%) & 54 (16\%) & 30 (9\%) & 155 (47\%) & 0 (0\%) \\
1.5 & 2.0 & 9.0 & 9.5 & 8985 & 530 (5\%) & 1870 (20\%) & 6057 (67\%) & 448 (4\%) & 80 (0\%) \\
1.5 & 2.0 & 9.5 & 10.0 & 5020 & 347 (6\%) & 1730 (34\%) & 2421 (48\%) & 488 (9\%) & 34 (0\%) \\
1.5 & 2.0 & 10.0 & 10.5 & 2620 & 518 (19\%) & 787 (30\%) & 939 (35\%) & 365 (13\%) & 11 (0\%) \\
1.5 & 2.0 & 10.5 & 11.0 & 1612 & 361 (22\%) & 457 (28\%) & 260 (16\%) & 530 (32\%) & 4 (0\%) \\
1.5 & 2.0 & 11.0 & 11.5 & 309 & 68 (22\%) & 74 (23\%) & 24 (7\%) & 143 (46\%) & 0 (0\%) \\
2.0 & 2.5 & 9.0 & 9.5 & 6280 & 305 (4\%) & 660 (10\%) & 5067 (80\%) & 177 (2\%) & 71 (1\%) \\
2.0 & 2.5 & 9.5 & 10.0 & 3512 & 176 (5\%) & 768 (21\%) & 2322 (66\%) & 205 (5\%) & 41 (1\%) \\
2.0 & 2.5 & 10.0 & 10.5 & 1894 & 316 (16\%) & 389 (20\%) & 1016 (53\%) & 151 (7\%) & 22 (1\%) \\
2.0 & 2.5 & 10.5 & 11.0 & 934 & 208 (22\%) & 169 (18\%) & 294 (31\%) & 255 (27\%) & 8 (0\%) \\
2.0 & 2.5 & 11.0 & 11.5 & 209 & 42 (20\%) & 51 (24\%) & 38 (18\%) & 78 (37\%) & 0 (0\%) \\
2.5 & 3.0 & 9.0 & 9.5 & 4202 & 201 (4\%) & 248 (5\%) & 3582 (85\%) & 72 (1\%) & 99 (2\%) \\
2.5 & 3.0 & 9.5 & 10.0 & 2368 & 126 (5\%) & 320 (13\%) & 1790 (75\%) & 87 (3\%) & 45 (1\%) \\
2.5 & 3.0 & 10.0 & 10.5 & 1220 & 186 (15\%) & 177 (14\%) & 775 (63\%) & 70 (5\%) & 12 (0\%) \\
2.5 & 3.0 & 10.5 & 11.0 & 509 & 120 (23\%) & 74 (14\%) & 197 (38\%) & 114 (22\%) & 4 (0\%) \\
2.5 & 3.0 & 11.0 & 11.5 & 106 & 20 (18\%) & 22 (20\%) & 30 (28\%) & 34 (32\%) & 0 (0\%) \\
3.0 & 3.5 & 9.0 & 9.5 & 3154 & 36 (1\%) & 41 (1\%) & 2933 (92\%) & 4 (0\%) & 140 (4\%) \\
3.0 & 3.5 & 9.5 & 10.0 & 1635 & 63 (3\%) & 127 (7\%) & 1384 (84\%) & 7 (0\%) & 54 (3\%) \\
3.0 & 3.5 & 10.0 & 10.5 & 912 & 110 (12\%) & 150 (16\%) & 558 (61\%) & 24 (2\%) & 70 (7\%) \\
3.0 & 3.5 & 10.5 & 11.0 & 473 & 86 (18\%) & 123 (26\%) & 178 (37\%) & 57 (12\%) & 29 (6\%) \\
3.0 & 3.5 & 11.0 & 11.5 & 131 & 10 (7\%) & 57 (43\%) & 22 (16\%) & 38 (29\%) & 4 (3\%) \\
3.5 & 4.5 & 9.0 & 9.5 & 3970 & 25 (0\%) & 13 (0\%) & 3472 (87\%) & 0 (0\%) & 460 (11\%) \\
3.5 & 4.5 & 9.5 & 10.0 & 1823 & 59 (3\%) & 77 (4\%) & 1524 (83\%) & 8 (0\%) & 155 (8\%) \\
3.5 & 4.5 & 10.0 & 10.5 & 873 & 67 (7\%) & 131 (15\%) & 499 (57\%) & 27 (3\%) & 149 (7\%) \\
3.5 & 4.5 & 10.5 & 11.0 & 310 & 36 (11\%) & 73 (23\%) & 94 (30\%) & 23 (7\%) & 84 (27\%) \\
3.5 & 4.5 & 11.0 & 11.5 & 60 & 2 (3\%) & 17 (28\%) & 11 (18\%) & 9 (15\%) & 21 (35\%) \\
4.5 & 5.5 & 9.0 & 9.5 & 2349 & 11 (0\%) & 4 (0\%) & 1972 (83\%) & 0 (0\%) & 362 (15\%) \\
4.5 & 5.5 & 9.5 & 10.0 & 790 & 13 (1\%) & 5 (0\%) & 682 (86\%) & 0 (0\%) & 90 (11\%) \\
4.5 & 5.5 & 10.0 & 10.5 & 201 & 14 (6\%) & 5 (2\%) & 110 (54\%) & 2 (0\%) & 70 (34\%) \\
4.5 & 5.5 & 10.5 & 11.0 & 109 & 7 (6\%) & 6 (5\%) & 19 (17\%) & 1 (0\%) & 76 (69\%) \\
4.5 & 5.5 & 11.0 & 11.5 & 21 & 0 (0\%) & 0 (0\%) & 2 (9\%) & 0 (0\%) &
19 (90\%) \\
5.5 & 6.5 & 9.0 & 9.5 & 972 & 1 (0\%) & 0 (0\%) & 814 (83\%) & 0 (0\%) & 157 (16\%) \\
5.5 & 6.5 & 9.5 & 10.0 & 290 & 1 (0\%) & 0 (0\%) & 254 (87\%) & 0 (0\%) & 35 (12\%) \\
5.5 & 6.5 & 10.0 & 10.5 & 58 & 1 (1\%) & 0 (0\%) & 44 (75\%) & 0 (0\%) & 12 (22\%) \\
5.5 & 6.5 & 10.5 & 11.0 & 28 & 2 (7\%) & 4 (14\%) & 7 (25\%) & 1 (3\%) & 14 (50\%) \\
5.5 & 6.5 & 11.0 & 11.5 & 14 & 2 (14\%) & 0 (0\%) & 3 (21\%) & 0 (0\%) & 9 (64\%) \\
6.5 & 7.5 & 9.0 & 9.5 & 667 & 2 (0\%) & 0 (0\%) & 428 (64\%) & 0 (0\%) & 237 (35\%) \\
6.5 & 7.5 & 9.5 & 10.0 & 198 & 6 (3\%) & 0 (0\%) & 144 (72\%) & 0 (0\%) & 48 (24\%) \\
6.5 & 7.5 & 10.0 & 10.5 & 35 & 1 (2\%) & 0 (0\%) & 21 (60\%) & 0 (0\%) & 13 (37\%) \\
6.5 & 7.5 & 10.5 & 11.0 & 14 & 1 (7\%) & 0 (0\%) & 5 (35\%) & 0 (0\%)&
8 (57\%) \\
6.5 & 7.5 & 11.0 & 11.5 & 9 & 0 (0\%) & 0 (0\%) & 0 (0\%) & 0 (0\%) &
9 (100\%) \\
\hline
	\end{tabular}
\begin{flushleft}
Note: a - The total is the total number of galaxies in this range of redshift
and stellar mass. The numbers in the other columns are the number of galaxies,
and in brackets the percentage of the total, in each morphological class. `uc' stands for unclassified.
\end{flushleft}
\end{table*}

\begin{table*}
	\centering
	\caption{Morphological Composition - continued}
	\label{tab:example_table}
	\begin{tabular}{llllllllll} 
		\hline
		$\rm z_{low}$ & $\rm z_{up}$ & $\rm log_{10}(M_{low}/M_{\odot})$ & $\rm log_{10}(M_{up}/M_{\odot})$ & Total$^a$ & spheroids$^a$ & disks$^a$ & irregulars$^a$ & bulge-dominated & uc$^a$ \\
		\hline
7.5 & 8.5 & 9.0 & 9.5 & 415 & 2 (0\%) & 0 (0\%) & 220 (53\%) & 0 (0\%) & 193 (46\%) \\
7.5 & 8.5 & 9.5 & 10.0 & 183 & 9 (4\%) & 0 (0\%) & 86 (46\%) & 0 (0\%) & 88 (48\%) \\
7.5 & 8.5 & 10.0 & 10.5 & 25 & 0 (0\%) & 0 (0\%) & 13 (52\%) & 0 (0\%) & 12 (48\%) \\
7.5 & 8.5 & 10.5 & 11.0 & 16 & 0 (0\%) & 1 (6\%) & 5 (31\%) & 0 (0\%) & 10 (62\%) \\
7.5 & 8.5 & 11.0 & 11.5 & 3 & 0 (0\%) & 0 (0\%) & 0 (0\%) & 0 (0\%) &
3 (100\%) \\
8.5 & 12.0 & 9.0 & 9.5 & 128 & 1 (0\%) & 0 (0\%) & 92 (71\%) & 0 (0\%) & 35 (27\%) \\
8.5 & 12.0 & 9.5 & 10.0 & 105 & 0 (0\%) & 0 (0\%) & 71 (67\%) & 0 (0\%) & 34 (32\%) \\
8.5 & 12.0 & 10.0 & 10.5 & 97 & 0 (0\%) & 0 (0\%) & 49 (50\%) & 1 (1\%) & 47 (48\%) \\
8.5 & 12.0 & 10.5 & 11.0 & 26 & 0 (0\%) & 0 (0\%) & 16 (61\%) & 0 (0\%) & 10 (38\%)\\
8.5 & 12.0 & 11.0 & 11.5 & 5 & 0 (0\%) & 0 (0\%) & 2 (40\%) & 0 (0\%) 
& 3 (60\%) \\
\hline
	\end{tabular}
\begin{flushleft}
Note: a - The total is the total number of galaxies in this range of redshift
and stellar mass. The numbers in the other columns are the number of galaxies,
and in brackets the percentage of the total, in each morphological class. `uc' stands for unclassified.
\end{flushleft}
\end{table*}

\section{Biases in the estimates of the temperature of dust in galaxies}

The dust in a real galaxy has a range of temperatures. The most useful
single number that describes the temperature of the dust is probably
the mean mass-weighted temperature of the dust, which is what is needed, for
example, for an accurate estimate of the mass of the dust.

There is a well-known problem that arises when one estimates the temperature
of the dust in a galaxy by fitting a single modified blackbody to its
far-infrared and submillimetre photometry. The problem
arises because warm dust radiates more strongly and at a shorter wavelength
than the same mass of cold dust. The effect is that the temperature
estimated from fitting the modified
blackbody is always higher than the mean mass-weighted temperature of the dust
\citep{eales1989}. For a given set of observation wavelengths, the difference
between the two temperatures - what one measures and what one needs - should
increase with redshift as the observations move, in the rest-frame of the galaxy,
to shorter wavelengths, where the effects of the warm dust are larger.
Several groups have investigated this bias using cosmological simulations
\citep{behrens2018,liang2019,lower2024,sommovigo2025}.
\citet{liang2019}, for example, found that although the mean mass-weighted temperature
of the dust in their simulations did not increase with redshift, the observed temperature of the dust did.

\begin{figure}
	\includegraphics[width=\columnwidth]{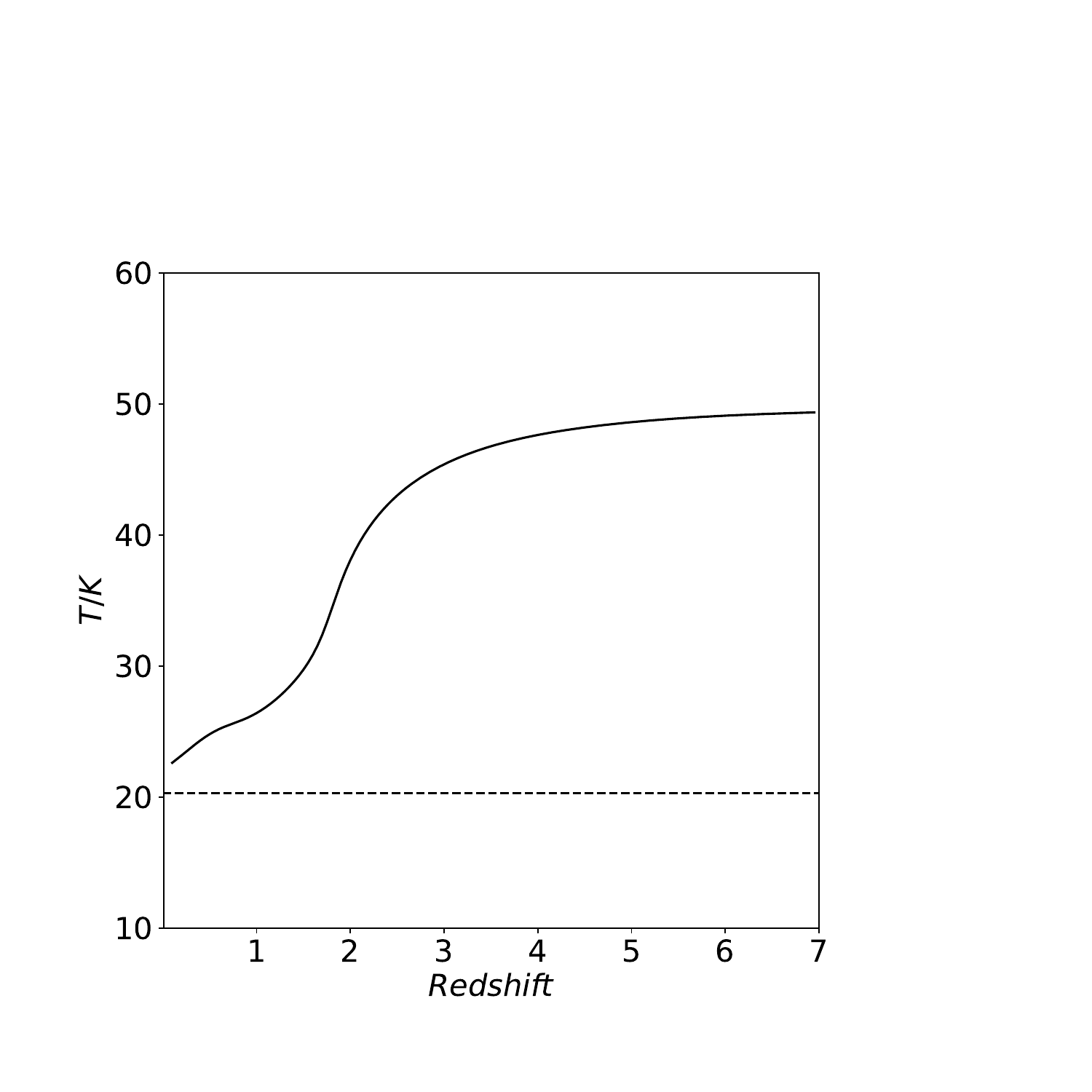}
	\caption{Observed dust temperature versus redshift found from a stacking
    analysis for a sample of galaxies in which 99\% of the galaxies
contain dust at 20 K and 1\% contain dust at 50 K. The horizontal dashed
line show the mean mass-weighted temperature of the dust.}
	\label{fig:morphologies}
\end{figure}

In this section, we show that even when all the
dust in individual galaxies is at the same temperature, the bias
is still present when the temperature is estimated from a 
stacking analysis. The most common way to estimate
the dust temperature in a stacking analysis is to measure the mean
flux densities of a sample of galaxies at a set of wavelengths
and then fit these with a modified blackbody \citep{magdis2012,magnelli2014,bethermin2015,schreiber2018,Bouwens2020,viero2022}.

We model this in the following way.
We assume that every galaxy contains the same mass of dust with a uniform
temperature, with 
99\% of the galaxies in the sample containing dust at 20 K and the remaining
1\% containing dust at 50 K. We assume that the
stacking is being carried out at 100, 160, 250, 350, 500 and 850 $\mu$m, which
are common wavelengths used
in stacking analyses. We assume that the sample of galaxies
is at a redshift $z$, calculate the mean flux densities of the sample
at each wavelength, and then fit the results
with a modified black body in the rest-frame of the
sample.
We assume that the value of the emissivity index, $\beta$, in the
modified blackbody is 2 and that the mean flux densities have 10\% errors.

Figure B1 shows the result. The mean mass-weighted dust temperature
is 20.3 K, but the measured dust temperature steadily climbs
with redshift, asymptotically approaching the temperature of the
warmer galaxies. The explanation for this is the
same as the explanation given above for the difference between the
observed temperature and the mean mass-weighted dust temperature
of individual galaxies.
As the redshift increases, the part of the
rest-frame SED sampled by the observations will move to shorter
wavelengths and the photometry is increasingly dominated
by the galaxies with the warmer temperature.

This is illustrated in Figure B2, which shows
the mean SED of a galaxy sample at $\rm z=3$ in our model 
and the relative contributions to it of the warm and cold galaxies. At
the shortest wavelengths, almost all the flux is
contributed by the warm galaxies, even though they only contain 1\%
of the mass of the dust. The difference between the observed dust temperature
and the mean mass-weighted dust temperature increases with redshift as the photometry
becomes increasingly dominated by the radiation from the warmer galaxies.
One of the most interesting results from a stacking analysis is the
recent claim that the dust in galaxies at $\rm z \sim 7$ has a temperature
of $\rm \simeq 100\ K$ \citep{viero2022}. Our analysis suggests that
the correct interpretation of this result is not that this is
the mean temperature of the dust at this redshift, but that some
galaxies at this redshift contain dust at this temperature, for
which there is evidence from observations of individual galaxies
\citep{bakx2020}.

\begin{figure}
	\includegraphics[width=\columnwidth]{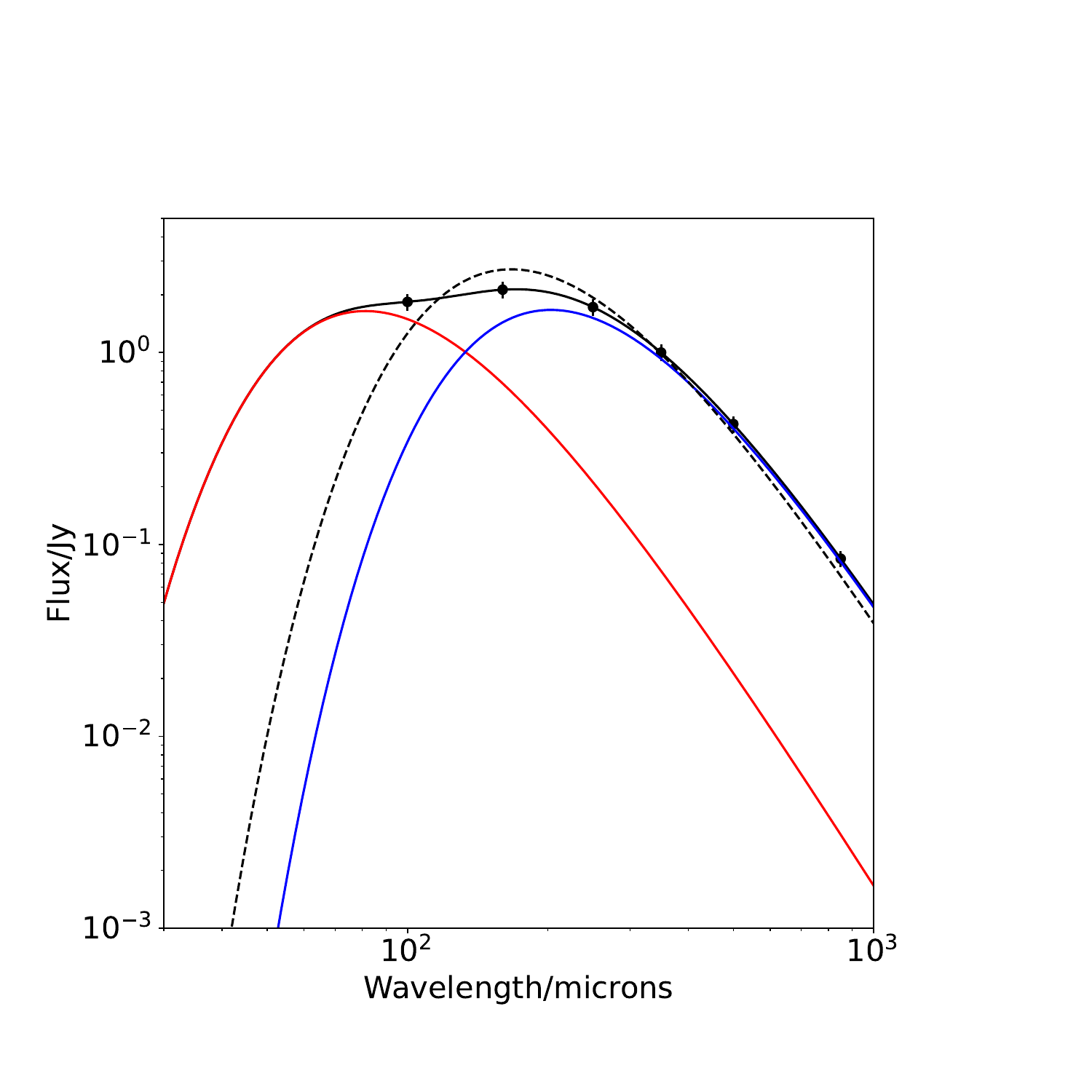}
	\caption{Spectral energy distribution (black continuous line) of a sample
of galaxies at z = 3 
for the model described in the text.
The red line shows the contribution from the warm ($\rm T=50\ K$) galaxies and the blue line the contribution
from the cold ($\rm T=20\ K$) galaxies. The dashed line shows
the fit of a modified blackbody at $\rm z =3$ to mean flux densities (the points and errors)
of this sample.}
	\label{fig:morphologies}
\end{figure}


\bsp	
\label{lastpage}
\end{document}